%% file: version_5.tex
\documentclass[a4paper,11pt]{article}
\usepackage{jheppub} 
\usepackage{lineno}
\usepackage{import}
\input{packdefs}
\usepackage{slashed}

\title{\boldmath dS/CFT Holography including Fermionic Fields}
\author[a]{Akashdeep Roy}
\affiliation[a]{\it Department of Theoretical Physics,
	Tata Institute of Fundamental Research,\\  Colaba, Mumbai, India, 400005}
\emailAdd{roy.akashdeep@tifr.res.in}

\preprint{TIFR/TH/26-25}

\vspace{1cm}
\abstract{We investigate some aspects of holography in de Sitter space, pertaining to the proposal for the dual theory residing on the late-time boundary \cite{NG_malda}, including a Fermionic field that satisfies the Dirac equation. The modes of the solutions near the boundary are found to possess equal and oscillatory fall-off, similar to the scalar field belonging to the principal series representation of the isometry group of de Sitter space. We evaluate, by suitably adding a boundary term to the Dirac action, the wavefunctional of the Bunch Davies vacuum in path integral formalism in terms of the projections of the field on the eigenspaces of the time-like ${\gamma}$-matrix. We validate our path-integral construction of the wavefunctional by reproducing the two-point function between the field operators in the vacuum state as computed in the canonical formalism.
By appropriately identifying the boundary sources in terms of the projected components of the bulk field, we find that a single Dirac field in the bulk corresponds to a pair of primary spinor fields in the boundary theory and show that the resulting wavefunction coefficients satisfy the conformal Ward identities. We check the robustness of our holographic map under the addition of a Yukawa interaction considering the two cases where the scalar field in the interaction term belongs to the complementary series and the principal series representations of the isometry group of de Sitter space respectively.

}

\begin{document}
	\maketitle
	\flushbottom
	
	
	\section{Introduction}
	\label{Sec.intro}
	
	de Sitter (dS) space, the maximally symmetric solution of Einstein's equation of gravity with positive cosmological constant, has gained considerable interest in the studies of quantum gravity. One captivating question is whether dS space has a holographic description. Being inspired from the concrete realisation of AdS/CFT \cite{Maldacena:1997re,Witten:1998qj,Aharony:1999ti,DeWolfe:2018dkl}, it is also suggested that dS space should carry a similar holographic correspondence, given by the equality of late-time wavefunctional in the bulk with the partition function of a CFT that lives on the space-like hypersurface located at future infinity ${\cal I}^+$ \cite{NG_malda}.     
	
	Although this version appears very similar to the GKPW dictionary in AdS/CFT \cite{Gubser:1998bc,Witten:1998qj}, 
	it is manifestly clear from further investigations \cite{Witten:2001kn, Strominger:2001pn, Goldar:2024crc, roy_dscft, Dey:2026lcu, Yang:2026lne} that dS/CFT differs in many aspects from AdS/CFT holography. The differences are even prominent in case of minimally-coupled scalar theory.
	For example, modes of the scalar fields of mass $M_0>{d\over2R}$ in dS${}_{d+1}$ with radius $R$, belonging to the principal series representation of dS isometry group, possess equal and oscillatory fall-off near the boundary ${\cal I}^+$ which has no analogue in AdS. For the Bunch Davies vacuum, the source in this case is identified in the coherent state basis representation of the wavefunctional \cite{roy_dscft}, also see \cite{Isono_dscft} for the explanation in terms of mixed boundary condition. Even in case of $M_0<{d\over2R}$, where the identification of the boundary source is quite analogous to AdS/CFT, the resulting correlation function has a cut-off dependent local term in position space which is important to retain as it ensures the gauge-invariance of the wavefunction satisfying the Wheeler de-Witt equation. Even if the finite term in the correlation function can be analytically continued from the corresponding AdS result, violation of reflection positivity of the term implies that one cannot continue the Euclidean boundary theory to its Lorentzian version with Hermitian operators etc. A brief discussion reviewing the source identifications for scalar fields in dS holography is given in Appendix \ref{app.scalar}.
	
	In this paper, we study some aspects of the late-time boundary version of dS/CFT correspondence for Bunch Davies (BD) vacuum state including Fermionic fields. It is essential for this study to construct the wavefunctional of this state. In scalar theory, wave functional of the BD vacuum, for example, can be understood from canonical quantisation, also from the path integral, in a way analogous to that for the ground state of bosonic harmonic oscillator. 
	In Fermionic field theory, we construct the wavefunctional of BD vacuum state via path integral. Compared to the bosonic case, this construction requires more care due to the following reasons. A bosonic theory has, in general, two derivatives in the action. For example, consider a free scalar theory
	\begin{equation}
		S_{sc}=\int d^{d+1}x\sqrt{-g}\del_\mu\phi\del^\mu\phi + \cdots
	\end{equation}
	where $\phi$ is the scalar field. The canonical momentum given by
	\begin{equation}
		\pi_\phi\sim \del_0\phi
	\end{equation}
	has a time derivative acting on the field.
	Therefore specification of $\phi$ at the time boundary $\Sigma$ of path integral via Dirichlet condition does not automatically specify $\pi_\phi$ at $\Sigma$. The wavefunctional of a state can be solely represented in field eigenstate basis in this case. On the other hand, the canonical momentum computed from the Dirac Lagrangian $L$ given in \eq{frdiract} results 
	\begin{equation}
		\label{mompsi}
		\pi_\psi={\delta L\over\delta (\del_0\psi)}= {iR^d\over2}(-\eta)^{-d}\psi^\dagger
	\end{equation}
	Therefore, specification of the field $\psi$ at $\Sigma$ requires the path integral to be written in terms of the variable $\psi$, and its adjoint $\psi^\dagger$ as an independent spinor field. This in turn requires independent specification of $\psi^\dagger$ on $\Sigma$ simultaneous to $\psi$ violating uncertainty principle \cite{Henneaux_ferads}. It is therefore clear that Dirichlet boundary condition is over-restrictive for Fermionic fields because the Dirac action \eq{frdiract} appears with single derivative operator. More fundamentally, this issue lies in the variational problem of Fermionic field theory as discussed in \cite{Chen_flatwfc}.    
	The key idea to resolve the issue is to express the wavefunctional in terms of the projections of the variables $\psi$ and $\psi^\dagger$ onto the eigenspaces of $\gamma^0$,  e.g. see \eq{psibypsi} for a particular representation. We refer \cite{Henneaux_ferads, Henning_ferads} for similar comments in AdS/CFT and \cite{Ageev_flatfer} in flat space holography.
	This wavefunctional is found carrying richer structure, thanks to the spinors, compared to that in Fermionic oscillator in quantum mechanics with spin-1/2 particle where the dynamics is best described using Grassmann variables and the ground state wavefunction is given just by a bosonic constant \cite{justin_piqm}.
	
	To draw the motivation, let us start our discussion with free Fermionic fields in AdS/CFT correspondence along the lines of \cite{Henneaux_ferads}. Metric of Euclidean AdS of radius $L$ in Poincare coordinates is given by
	\begin{equation}
		ds^2={L^2\over z^2}(dz^2+d{\bf x}^2)
	\end{equation} 
	Dirac equation for Fermionic field $\psi$ of mass $M$ in this space is given by
	\begin{equation}
		\label{eomads}
		(z\Gamma^a\delta^\mu_a\del_\mu-{d\over2}\Gamma^0-ML)\psi=0 ~~,~~\bar{\psi}(\overleftarrow{\del}_\mu z\Gamma^a\delta^\mu_a-{d\over2}\Gamma^0+ML)=0
	\end{equation}
	where $\bar{\psi}$ is the conjugate spinor of $\psi$ and
	$\Gamma^a$ is the element of Clifford algebra in flat Euclidean space\footnote{The index $0$ labels the radial/holographic tangent direction.}.
	The solutions of the Dirac equation in this space are classified in two types -- the first type that is annihilated by $1+\Gamma^0$ to the leading order is given asymptotically ($z\rightarrow0$) by
	\begin{equation}
		\lim_{z\rightarrow0}\psi_-({\bf x},z)= z^{{d\over2}-ML}{\psi}_0({\bf x})
	\end{equation} 
	where $\psi_0({\bf x})$ is arbitrary spinor satisfying $\Gamma^0 {\psi}_0=-{\psi}_0$; and the second type that is annihilated by $1-\Gamma^0$ to the leading order is given asymptotically ($z\rightarrow0$) by
	\begin{equation}
		\lim_{z\rightarrow0}\psi_+({\bf x},z)= z^{{d\over2}+ML}{\chi}_0({\bf x})
	\end{equation}
	where $\chi_0({\bf x})$ is another arbitrary spinor satisfying $\Gamma^0 {\chi}_0={\chi}_0$.
	The general solution, given by the linear superposition of $\psi_-$ and $\psi_+$, is therefore dominated by ${\psi}_0$ asymptotically as $z\rightarrow0$. The partition function $Z_{AdS}$ in the bulk is computed in the semi-classical limit by fixing ${\psi}_0$ and $\bar{\psi}_0$ on the boundary $z=0$. Considering $\{{\psi}_0,\bar{\psi}_0\}$ as the source of the boundary theory, $Z_{AdS}$ has been shown taking the form of a CFT partition function. The corresponding two point correlation function is in the form of a CFT two point correlator with operators of conformal dimension
	\begin{equation}
		\label{cdimads}
		\Delta={d\over2}+ML
	\end{equation}     
	On the other hand, working in Poincare coordinates in which the metric of dS space takes the form,
	\begin{equation}
		\label{poinds}
		ds^2={R^2\over \eta^2}(-d\eta^2+d{\bf x}^2)
	\end{equation}
	the free Fermionic modes near the boundary $\eta=0$ are given in the expanding Poincare patch ($\eta<0$) by
	\begin{equation}
		\psi({\bf x},\eta)=(-\eta)^{{d\over2}+iMR}\psi_+({\bf x})+(-\eta)^{{d\over2}-iMR}\psi_-({\bf x})
	\end{equation}	
	where $\psi_+,\psi_-$ are arbitrary spinors satisfying $\gamma^0\psi_\pm=\pm \psi_\pm$, $\gamma^a$ being the element of clifford algebra in Minkowski space. We refer Section \ref{sec.clsol} for more details. It is thereby to note that, the modes, in contrast to those in AdS space, possess equal and oscillatory fall-off near the boundary. A similar situation is encountered in case of scalar field with mass $M_0>{d\over2R}$ in dS space, as we have discussed above. Interestingly, in case of Fermions in dS space, it extends for all non-zero $M$ values. 
	
	Nonetheless, we show in this paper that, for Bunch-Davies vacuum state, one can identify the sources \eq{rhopm_s}, very similar to that in AdS/CFT, in terms of $\{\psi_+,\psi_-\}$, the projections of the field on the eigenspaces of $\gamma^0$. Expressing the wavefunctional, represented as a functional of $(\psi_+^\dagger,\psi_-)$, in terms of these sources as \eq{latewfrho}, the corresponding wavefunction coefficient takes the form of CFT two point function in free theory, \eq{FFpos}. This correlation function involves two boundary fields $\{{\hat\zeta}_+,{\hat\zeta}_-\}$ of dimensions
	\begin{equation}
		\Delta_+={d\over2}+iMR ~~,~~ \Delta_-={d\over2}-iMR
	\end{equation} 
	respectively. We emphasize the fact that the wavefunction can also be represented as a functional of $\{\psi_-^\dagger,\psi_+\}$, \eq{psibypsial} which alters the role of Fermion and anti-Fermion compared to the representation we mentioned above in $\{\psi_+^\dagger,\psi_-\}$, \eq{psibypsi}, as $\psi_+$ and $\psi_-$ belong to the eigenspaces of $\gamma^0$ with eigenvalue $+1$ and $-1$ respectively. The corresponding wavefunction coefficient takes the form of CFT two point function, \eq{wfcal}, involving the same boundary fields $\{{\hat\zeta}_+,{\hat\zeta}_-\}$.
	We also show that the invariance of the wavefunctional under spatial and time reparametrisations gives rise to constraints on the wavefunction coefficient in the form of CFT Ward identities. We find our identifications robust under the inclusion of Yukawa potential\footnote{See \cite{Sui:2025him} for discussions regarding the observational implications of adding Yukawa interaction in inflationary cosmology.} considering both overdamped and underdamped scalars.

	%
	
	
	Rest of the contents of the paper are organised as follows. In section \ref{sec.clsol}, we determine the solution of Dirac equation in dS space in $d+1$ dimensions. In section \ref{sec.bdwf}, we evaluate wavefunctional of the Bunch Davies vacuum for free Fermionic fields in path integral formalism. In section \ref{sec.holo}, we identify the sources in the boundary theory in terms of the projections of bulk Fermionic fields such that the resulting wavefunction coefficients, at late-time limit, take the form of a CFT correlation function, and discuss some subtleties arising in this context. In section \ref{Sec.WI}, we conclude, from the invariance of the wavefunction under spatial and time reparametrisation, that the wavefunction coefficients satisfy corresponding CFT Ward identities. We summarise our results and put some remarks in section \ref{Sec.discuss}.  Supplementary calculations are provided in appendices \ref{App.corwfc}--\ref{App.int}.

	\paragraph{Notations and Conventions} 
	In the rest of the paper, we denote space-time components in the tangent space by indices $\{a,b,\cdots\} \in \{0,1,\cdots,d\}$ and those in the bulk manifold by greek letters $\{\mu,\nu,\cdots\} \in \{0,1,\cdots,d\}$. The spatial indices in the bulk are particularly denoted by $\{i,j,\cdots\}\in \{1,\cdots,d\}$. `Structural components', by which we mean e.g. matrix elements of $\gamma^a$, spinor components of $\psi$ etc are denoted with indices $\{{\dot\alpha},{\dot\beta},\cdots\}$. \\  
	We fix the sign conventions as follows. We consider the metric with mostly positive signature along with 
	\begin{equation}
		\label{gacomdag}
		\brs{\gamma^a,\gamma^b}=-2\eta^{ab}{\mathbb I}~~~{\rm and}~~~ (\gamma^a)^\dagger=\gamma^0\gamma^a\gamma^0
	\end{equation}
	which implies the following properties of $\gamma$-matrices
	\begin{equation}
		\label{gsqdag}
		(\gamma^0)^2={\mathbb I}~~~~(\gamma^i)^2=-{\mathbb I}~~;~~(\gamma^0)^\dagger={\gamma^0}~~~~(\gamma^i)^\dagger=-{\gamma^i}
	\end{equation}
	To note, $\gamma$-matrices are always defined in the tangent space as $\gamma^a$. Wherever we write $\gamma^\mu$, it is to be considered as $\gamma^a\delta_a^\mu$. We work in natural units $G={\hbar}=c=1$. 
	\section{Classical Solution of Dirac Equation}
	\label{sec.clsol}
	Theory of Fermionic fields, or more general spinor fields, on curved spacetime traditionally relies on n-bein formalism\footnote{The name `n-bein' stands for referring $n$ space-time dimensions. We are using the name more generally to any dimensions. See \cite{ferwover} for constructing Fermionic fields on curved spaces without using n-bein.}. The formalism introduces at each point on the manifold a local orthonormal frame which has a basis one-form $\omega^a$. Introducing $\{dx^\mu\}$ as a set of coordinate basis one-forms for the cotangent space, n-bein $e^\mu_a$ is defined as
	\begin{equation}
		\label{e1form}
		\omega^a=e_\mu^a(x) dx^\mu
	\end{equation}
	In terms of n-bein, the metric of the manifold can be related to the metric of the co-tangent space at point $x$ as
	\begin{equation}
		\label{gtoeta}
		g_{\mu\nu}(x)=e^a_\mu(x)e^b_\nu(x)\eta_{ab}
	\end{equation}
	n-bein satisfies the following inversion formulae,
	\begin{equation}
		\label{einvform}
		e^a_\mu(x) e_a^\nu(x)=\delta^\nu_\mu ~~,~~ e^b_\nu(x) e_a^\nu(x)=\delta^b_a
	\end{equation}
	by which, we can invert \eq{gtoeta} to get
	\begin{equation}
		\label{etatog}
		\eta_{ab}=e_a^\mu(x)e_b^\nu(x)g_{\mu\nu}(x)
	\end{equation}
	See \cite{toms-parker} for more discussions.
	
	We consider the Dirac action on curved space-time for fields with mass $M$ given by
	\begin{equation}
		\label{frdiract}
		S_D[\bar{\psi},\psi]={1\over2}\int d^{d+1}x\sqrt{-g}\brt{{i}\brf{\bar{\psi}e^\mu_a\gamma^a\nabla_\mu\psi}-M\bar{\psi}\psi}
	\end{equation}
	where
	\begin{equation}
		\label{barpsidef}
		\bar{\psi}=\psi^\dagger\gamma^0
	\end{equation}  
	Extremisation of the action w.r.t. $\bar\psi$ and $\psi$ produce equation of motion (EoM) as
	\begin{equation}
		\label{freomdirac}
		[ie^\mu_a\gamma^a\nabla_\mu-M]\psi=0
	\end{equation}
	and,
	\begin{equation}
		\label{freomdiracb}
		{\bar\psi}[i\overleftarrow{\nabla}_\mu e^\mu_a\gamma^a+M]=0
	\end{equation}
	respectively. The action of $\nabla_\mu$ on Dirac field $\psi(x^\mu), {\bar\psi}(x^\mu)$ is given by 
	\begin{equation}
		\label{covDfermion}
		\nabla_\mu\psi=\del_\mu\psi-{1\over8}\omega_\mu^{~ab}[\gamma_a,\gamma_b]\psi ~~,~~ {\bar\psi}\overleftarrow{\nabla}_\mu={\bar\psi}\brf{\overleftarrow{\del}_\mu+{1\over8}\omega_\mu^{~ab}[\gamma_a,\gamma_b]}
	\end{equation}
	where $\omega_\mu^{~ab}$ is components of spin connection given by
	\begin{equation}
		\label{spincon}
		\omega_{\mu~b}^{~a}=-e^{\nu}_b(\del_\mu e^a_\nu - \Gamma^\lambda_{\mu\nu}e^a_\lambda)
	\end{equation}
	One can understand from \eq{freomdirac} and \eq{freomdiracb} that at on-shell, $\psi^\dagger$ being the Hermitian adjoint of $\psi$.
	 
\begin{figure}[h]
		\centering
		\begin{tikzpicture}
			\draw (6,3) -- (6,0) -- (9,0) -- (9,3) -- (6,3) -- node[left]{$\mathcal{H}$}(9,0);
			\node[text width=3cm] at (9,3.3) {$\mathcal{I}^+$};
			\node[text width=3cm] at (9.5,2) {$\mathcal{R}_+$};
			\node[text width=3cm] at (8.5,0.6) {$\mathcal{R}_-$};
		\end{tikzpicture}
		\caption{Penrose diagram for de Sitter space.}    \label{poinpenrose}
	\end{figure}
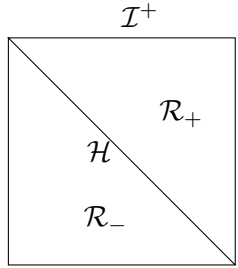

	In de Sitter space, the metric in Poincare coordinates is given in \eq{poinds} 
	where $R$ is the radius of the dS space. We consider upper (expanding) Poincare patch  ${\cal R}_+$, identified by the range of $\eta\in(-\infty,0)$ where $\eta=-\infty$ is the Poincare horizon $\cal H$ and $\eta=0$ is the future spacelike boundary ${\cal I}^+$, see Fig. \ref{poinpenrose}. Also note that the lower (contracting) Poincare patch ${\cal R}_-$ is identified by the range $\eta\in(\infty,0)$ where $\eta=\infty$ and $\eta=0$ are the locations of ${\cal H}$ and past spacelike boundary ${\cal I}^-$ respectively. 
	According to definition of n-bein, see \eq{gtoeta},
	\begin{equation}
		\label{nbeinds}
		e^a_\mu=\brf{{R\over(-\eta)}}\delta_\mu^a \implies e^\mu_a=\brf{{(-\eta)\over R}}\delta^\mu_a
	\end{equation}
	Given the metric \eq{poinds}, Christoffel symbol and spin connection \eq{spincon} is evaluated respectively as
	\begin{equation}
		\label{cristds}
		\Gamma^\lambda_{\mu\nu}=\brf{-{1\over\eta}}\brt{\delta^\lambda_\nu\delta^0_\mu + \delta^\lambda_\mu\delta^0_\nu - \eta^{\lambda 0}\eta_{\mu\nu}}
	\end{equation}
	and
	\begin{equation}
		\label{spinconds}
		\omega_\mu^{~ab}={1\over\eta}\brf{\eta^{a0}\delta^b_\mu -\eta^{b0}\delta^a_\mu}
	\end{equation}
	Hence from \eq{freomdirac}, we get by using \eq{gacomdag}
	\begin{equation}
		\label{freomdirds}
		\brt{i\eta\gamma^\mu\del_\mu-{id\over2}\gamma^0+MR}\psi=0
	\end{equation}
	Note that the equation of motion above can be obtained from that in EAdS space, \eq{eomads}, by the following analytic continuation
	\begin{equation}
		\label{analcont}
		L= iR~~,~~z=-i\eta ~~,~~ \Gamma^i= i\gamma^i
	\end{equation}
	\paragraph{Solution} Eq.\eqref{freomdirds} is a first order differential equation. To solve for the Fermionic field, we first make the equation second order differential by operating $\gamma^\mu\del_\mu$ on LHS. This gives
	\begin{equation}
		\brt{-i\eta\del^2-id\del_0+\brf{i\gamma^0+{id\over2}\gamma^0+MR}\gamma^\mu\del_\mu}\psi=0
	\end{equation}
	Using \eq{freomdirds} again, we get
	\begin{equation}
		\label{2ndorddif}
		\brt{\eta^2\del^2+d\eta\del_0+(-M^2R^2-{d^2\over4}-{d\over2}-iMR\gamma^0)}\psi=0
	\end{equation}
	To solve \eq{2ndorddif}, We decompose $\psi$ into its Fourier modes corresponding to Killing spatial coordinates
	\begin{equation}
		\label{psik}
		\psi({\bf x},\eta)=\fint{}\psi({\bf k},\eta)e^{i{\bf k}.{\bf x}}
	\end{equation}
	Putting above expression in \eq{2ndorddif}, we get
	\begin{equation}
		\label{2nddifk}
		\brt{-\eta^2\del_0^2+d\eta\del_0-(k^2\eta^2+M^2R^2+{d^2\over4}+{d\over2}+iMR\gamma^0)}\psi({\bf k},\eta)=0
	\end{equation}
	Note that $\gamma^0$ has degenerate eigenvalues $+1,-1$ and the degeneracy depends on no. of dimensions of the space-time. The corresponding eigenvectors $a_-,a_+$ given by
	\begin{equation}
		\label{g0eig}
		\gamma^0a_-^s({\bf k})=a^s_-({\bf k}) ~~,~~ \gamma^0a_+^s({\bf k})=-a_+^s({\bf k})~~,~~s=1,\cdots,2^{\brt{d+1\over2}-1}
	\end{equation}   
	forms a complete basis where $s$ is the index referring to degeneracies. The function $[x]$ determines the nearest integer less than or equal to $x$. The orthonormalisation of the eigenvectors are given by \footnote{We illustrate different representations of Gamma metrices and the spinor basis $a_\pm^s({\bf k})$ in appendix \ref{App.subtled}.}
	\begin{equation}
		\label{ortnorm}
		\sum_s (a_-^s)^\dagger{a_-^s}=1~~,~~\sum_s (a_+^s)^\dagger{a_+^s}=1~~,~~ (a_+^s)^\dagger{a_-^{s'}}=0~~,~~(a_\pm^s)^\dagger{a_\pm^{s'}}|_{s\neq s'}=0  
	\end{equation}
	We thereby expand $\psi({\bf k},\eta)$ in this basis most generally as
	\begin{equation}
		\label{psiexpand}
		\psi({\bf k},\eta)=\sum_{s}a_-^s({\bf k})f^s({\bf k},\eta)+a_+^s({\bf k})g^s({\bf k},\eta)
	\end{equation}
	Putting \eq{psiexpand} in \eq{2nddifk}, we get two 2nd order differential equations
	\begin{align}
		\label{two2nddif:f}
		&\brt{-\eta^2\del_0^2+d\eta\del_0-(k^2\eta^2+M^2R^2+{d^2\over4}+{d\over2}+iMR)}f({\bf k},\eta)=0\\
		&\brt{-\eta^2\del_0^2+d\eta\del_0-(k^2\eta^2+M^2R^2+{d^2\over4}+{d\over2}-iMR)}g({\bf k},\eta)=0 \label{two2nddif:g}
	\end{align}
	Solutions of the above equations are given as
	\begin{align}
		&f^s({\bf k},\eta)=C_1^s({\bf k})(-\eta)^{1+d\over2}H_{\nu_-}^1(-k\eta)+C_2^s({\bf k})(-\eta)^{1+d\over2}H_{\nu_-}^2(-k\eta) \label{fsol}\\
		&g^s({\bf k},\eta)=C_3^s({\bf k})(-\eta)^{1+d\over2}H_{\nu_+}^1(-k\eta)+C_4^s({\bf k})(-\eta)^{1+d\over2}H_{\nu_+}^2(-k\eta)\label{gsol}
	\end{align}
	where $C_1,\cdots,C_4$ are undetermined functions of ${\bf k}$, $k$ being its magnitude, and
	\begin{equation}
		\label{nupm}
		\nu_\pm={1\over2}\pm iMR
	\end{equation}
	Identifying the positive and negative frequency solution as
	\begin{align}
		&u_s({\bf k},\eta)\equiv a_-^s({\bf k}){C_1^s({\bf k})\over C_3^s({\bf k})}(-\eta)^{1+d\over2}H_{\nu_-}^1(-k\eta) + a_+^s({\bf k})(-\eta)^{1+d\over2}H_{\nu_+}^1(-k\eta)\label{u}\\
		&v_s({\bf k},\eta)\equiv a_-^s({\bf k}){C_2^s({\bf k})\over C_4^s({\bf k})}(-\eta)^{1+d\over2}H_{\nu_-}^2(-k\eta) + a_+^s({\bf k})(-\eta)^{1+d\over2}H_{\nu_+}^2(-k\eta)\label{v}
	\end{align}
	\eq{psiexpand} can be written as
	\begin{equation}
		\label{psisol2dif}
		\psi({\bf k},\eta)=\sum_s C_3^s({\bf k})u_s({\bf k},\eta)+C_4^s({\bf k})v_s({\bf k},\eta)
	\end{equation}
	Now, given \eq{freomdirds}, we find that $\psi({\bf k},\eta)$, defined in \eq{psik}, satisfies
	\begin{equation}
		\label{psi1difk}
		\brt{ i\eta\gamma^0\del_0-\eta\gamma^i k_i - {id\over2}\gamma^0 +MR } \psi({\bf k},\eta)=0
	\end{equation}
	Putting \eq{psisol2dif} in \eq{psi1difk} and demanding that $u$ and $v$, given in \eq{u}, \eqref{v}, should satisfy \eq{psi1difk} individually, we relate
	\begin{align}
		\label{a+a-rel}
		&a_-^s({\bf k})={\gamma^ik_i\over k}a_+^s({\bf k})
	\end{align}
	by the choice $C_1^s=-ie^{i\pi\nu_-}C_3^s$ and $C_2^s=-ie^{-i\pi\nu_-}C_4^s$. 
	Thereby we get the solution for $u,v$ as
	\begin{align}
		&u_s({\bf k},\eta)=\brt{-ie^{i\pi\nu_-}H^1_{\nu_-}(-k\eta)-{\gamma^i k_i\over k}H_{\nu_+}^1(-k\eta)}(-\eta)^{1+d\over2}a_-^s({\bf k})\label{usol}\\
		&v_s({\bf k},\eta)=\brt{-ie^{-i\pi\nu_-}{\gamma^i k_i\over k}H^2_{\nu_-}(-k\eta)+H_{\nu_+}^2(-k\eta)}(-\eta)^{1+d\over2}a_+^s({\bf k})\label{vsol}
	\end{align}
	Note that, by expanding the solutions above near $\eta=0$, we get by \eq{psisol2dif}
	\begin{equation}
		\lim_{\eta\rightarrow0} \psi({\bf k},\eta)=\sum_s
		(-\eta)^{{d\over2}+iMR}(\tilde{C}_1^s({\bf k}) a_-^s({\bf k}))+(-\eta)^{{d\over2}-iMR}(\tilde{C}_2^s({\bf k}) a_+^s({\bf k}))
	\end{equation}
	where
	\begin{equation}
		\tilde{C}_1^s({\bf k})=-{i2^{\nu_-}\Gamma[\nu_-]\over \pi}[C_3^s({\bf k})e^{\pi M R}+C_4^s({\bf k})e^{-\pi M R}]k^{-\nu_-} ~~,~~ \tilde{C}_2^s({\bf k})=-{i2^{\nu_+}\Gamma[\nu_+]\over \pi}[C_3^s({\bf k})-C_4^s({\bf k})]k^{-\nu_+}
	\end{equation}
	The above expansion shows that the modes of free Fermionic field possess equal and oscillatory fall-off near the boundary $\eta=0$ in dS space, as mentioned in Section \ref{Sec.intro}.  
	
	To determine the coefficients $C_3$ and $C_4$, one need to appropriately specify boundary conditions. It is convenient for this purpose to write the solution \eq{psisol2dif} projecting in the eigenspaces of $\gamma^0$.
	More precisely, $\psi$ can be decomposed into its projections as follows
	\begin{equation}
		\label{splitpsi}
		\psi=\psi_++\psi_-~;~~ \psi_\pm\equiv P_\pm\psi
	\end{equation}
	where $P_\pm$ is projection operator given by
	\begin{equation}
		\label{projop}
		P_\pm={1\pm\gamma^0\over2}
	\end{equation}
	{It is important to note that specification of either $\psi_+$ or $\psi_-$ at $\Sigma$ is not equivalent to specify $\psi$ at $\Sigma$ as $P_+,P_-$ are non-invertible.} Other properties of these projection operators are given by
	\begin{equation}
		\label{propPpm}
		P_\pm^2=P_\pm ~~~P_\pm^\dagger=P_\pm~~~P_+P_-=0
	\end{equation}
	Since $P_\pm$ is idempotent, \eq{splitpsi} implies
	\begin{equation}
		\label{Peig}
		P_\pm\psi_\pm=\psi_\pm
	\end{equation}
	which, by \eq{projop}, results
	\begin{equation}
		\label{psipmaseig}
		\gamma^0\psi_\pm=\pm\psi_\pm
	\end{equation}
	For later discussion, we also note here using \eq{splitpsi}, \eq{psipmaseig} and \eq{barpsidef} that
	\begin{equation}
	\label{barpsisplit}
		{\bar\psi}=\psi_+^\dagger-\psi_-^\dagger
	\end{equation}
	By \eq{psipmaseig}, one can write from \eq{psisol2dif}, introducing subscript $c$ to denote the classical value,
	\begin{align}
		&\psi_{+c}=-i(-\eta)^{1+d\over2}\sum_s C_3^s({\bf k})e^{i\pi\nu_-}H^1_{\nu_-}(-k\eta)a_-^s({\bf k})+C_4^s({\bf k})e^{-i\pi\nu_-}{\gamma^i k_i\over k}H^2_{\nu_-}(-k\eta)a_+^s({\bf k})\label{psi+c}\\
		&\psi_{-c}=(-\eta)^{1+d\over2}\sum_s -C_3^s({\bf k}){\gamma^i k_i\over k}H_{\nu_+}^1(-k\eta)a_-^s({\bf k})+C_4^s({\bf k})H_{\nu_+}^2(-k\eta)a_+^s({\bf k})\label{psi-c}
	\end{align}
	where we use \eq{usol} and \eqref{vsol}.  Conjugating the above two expressions, we get
	\begin{align}
		&(\psi_{+c})^\dagger=i(-\eta)^{1+d\over2}\sum_s C_3^s({\bf k})^*e^{-i\pi\nu_+}H^2_{\nu_+}(-k\eta)a_-^s({\bf k})^\dagger-a_+^s({\bf k})^\dagger C_4^s({\bf k})^*e^{i\pi\nu_+}{\gamma^i k_i\over k}H^1_{\nu_+}(-k\eta)\label{psi+cd}\\
		&(\psi_{-c})^\dagger=(-\eta)^{1+d\over2}\sum_s a_-^s({\bf k})^\dagger C_3^s({\bf k})^*{\gamma^i k_i\over k}H_{\nu_-}^2(-k\eta)+C_4^s({\bf k})^*H_{\nu_-}^1(-k\eta)a_+^s({\bf k})^\dagger \label{psi-cd}
	\end{align}
	
	It is thereby apparent that $C_3, C_4$ can be completely determined by the following mutually exclusive boundary conditions \footnote{Since classical solution of $\psi_+^\dagger$ is complex adjoint of $\psi_+$ in the bulk, the boundary value of $(\psi_{+c})^\dagger$ cannot be specified as independent data from $\psi_{+c}$. Similar argument holds for $\psi_{-c}$.}
	\begin{itemize}
		\item specifying $\psi_+,\psi_-$ independently at any finite time instant $\eta$.
		\item specifying $\psi_+,\psi_-$ independently at different time instant $\eta_1,\eta_2$ respectively.
		\item specifying $\psi_+,\psi_+^\dagger$ independently at different time instant $\eta_1,\eta_2$ respectively.
		\item specifying $\psi_-,\psi_-^\dagger$ independently at different time instant $\eta_1,\eta_2$ respectively.
	\end{itemize}
	Note that the first case is equivalent to specifying boundary condition on $\psi$ which is argued to determine the classical solution entirely. It is expected from the fact that Dirac equation is first order differential.

	\section{Wave Functional of Bunch Davies Vacuum}
	\label{sec.bdwf}
	In general, Bunch Davies vacuum is identified as one of the de Sitter invariant vacua in which the short distance singularity of correlators between operators behave as that in case of Minkowski vacuum and there is no non-local singularity. Discussions on the Bunch-Davies vacuum in literature can be found for scalar fields in \cite{Mottola:1984ar, bruceallen,BMS_states} and for higher integer spins in \cite{Allen:1986ta, Higuchi:1986py}. In case of Dirac fields, one can obtain the correlators in Bunch Davies vacuum, denoted by $\ket{0}$ later on, through canonically quantising the field as
	\begin{equation}
		{\psi}({\bf k},\eta)=\sum_s {\hat b}^s_{\bf k} N_u(k)u_s({\bf k},\eta)+({\hat c}^s_{\bf k})^{\dagger} N_v(k)v_s({\bf k},\eta) 
	\end{equation}  
	where ${\hat b},{\hat c}$ annihilate Bunch Davies vacuum
	\begin{equation}
		{\hat b}_{\bf k}^s\ket{0}=0~~,~~{\hat c}_{\bf k}^s\ket{0}=0
	\end{equation}
	and $u,v$ are given in \eq{usol}, \eqref{vsol} respectively. $N_u, N_v$, the normalisation constants fixed by the quantisation rules \eq{qrule} for Fermionic fields, are given by
	\begin{equation}
			N_u(k)=\sqrt{{\pi k}\over 4R^de^{\pi M R}}~~,~~ N_v(k)=\sqrt{{\pi k}\over 4R^de^{-\pi M R}}
	\end{equation}
	The two point correlator, for example, is computed in this vacuum as
	\begin{align}
		&\bra{0}{\hat \psi}({\bf k},\eta)\otimes{\hat\psi}^\dagger({\bf k}',\eta)\ket{0}={{\pi  k}\over 4R^d e^{\pi M R}}\sum_s u_s({\bf k},\eta)u_s^\dagger({\bf k},\eta)\delta({\bf k}-{\bf k}')\label{corrdagr}\\
		&\bra{0}{\hat\psi}^\dagger({\bf k},\eta)\otimes{\hat\psi}({\bf k}',\eta)\ket{0}={{\pi  k}\over 4R^d e^{-\pi M R}}\sum_s v_s({\bf k},\eta)v_s^\dagger({\bf k},\eta)\delta({\bf k}-{\bf k}')\label{corrdagl}
	\end{align}
	where $\otimes$ denotes the outer product of two spinors.
	See \cite{Cotaescu:2001cv, Anninos_dsfer, Cotaescu:2020gae, Fujikura:2026ucy} for related discussions in case of Dirac as well as Rarita-Schwinger fields.
	
	It is not understood well how to obtain wave functional of a state in Fermionic field theory from canonical quantisation, to the best of our knowledge. However, one can expect that the correlators obtained through canonical formalism in the Bunch Davies vacuum, \eq{corrdagr} and \eq{corrdagl}, can also be reproduced by computing expectation value of the insertions with respect to the wave functional of the vacuum state.
	In this section, we compute wave functional of Bunch Davies vacuum in path integral approach, and provide a check in appendix \ref{App.corwfc} that one can indeed obtain the correlators in free theory by computing expectation value with respect to the resulting wave functional.
	
	 Path integral for Fermionic field is given by
	\begin{equation}
		\label{pifer}
		PI=\int {\cal D}\psi {\cal D}\bar{\psi} e^{iS[\psi,\bar{\psi}]}
	\end{equation}
	where ${\bar\psi}$ and $\psi$ are treated as independent Grassmann fields. We evaluate this integral at the saddle point which takes the classical values of the fields.
	The BD vacuum in the path integral \eq{pifer} is determined by the condition that solutions of $\psi,{\bar\psi}$ independently vanish at the Poincare horizon $\cal H$, $\eta=-\infty$ by adding a small imaginary term to $\eta$. More precisely, we choose the solution of $\psi$ and ${\bar \psi}$ which vanishes at ${\cal H}$ for $\eta\rightarrow\eta(1-i\epsilon)$ and $\eta\rightarrow \eta(1+i\epsilon)$ respectively. 
		Near $\cal H$, we note that $H^2_\nu(-k\eta)$ vanishes under the prescription $\eta\rightarrow\eta(1-i\epsilon)$ with $\epsilon$ being small positive number as
	\begin{equation}
		\lim_{\eta\rightarrow-\infty} H^2_\nu(-k\eta)|_{\eta(1-i\epsilon)}\sim {e^{ik\eta(1-i\epsilon)}\over \sqrt{-k\eta}}\sim e^{\epsilon k \eta+ik\eta}~~,~~{\forall\nu} \label{h2hor}
	\end{equation}
	while  $H^1_\nu(-k\eta)$ blows up as
	\begin{equation}
		\lim_{\eta\rightarrow-\infty} H^1_\nu(-k\eta)|_{\eta(1-i\epsilon)}\sim {e^{-ik\eta(1-i\epsilon)}\over \sqrt{-k\eta}}\sim e^{-\epsilon k \eta-ik\eta}~~,~~{\forall\nu}
	\end{equation}
	Similarly, under $\eta\rightarrow \eta(1+i\epsilon)$, $H^1_\nu(-k\eta)$ vanishes while $H^2_\nu(-k\eta)$ diverges $\forall \nu$.
	Therefore by the vanishing condition of the field configuration at ${\cal H}$, we set $C_3^s({\bf k})=0$ in \eq{psi+c} and in \eq{psi-c} resulting
	\begin{align}
		&\psi_{+c}=-i(-\eta)^{1+d\over2}\sum_s C_4^s({\bf k})e^{-i\pi\nu_-}{\gamma^i k_i\over k}H^2_{\nu_-}(-k\eta)a_+^s({\bf k})\label{psi+cbd} \equiv \sum_s \psi_{+c}^s\\
		&\psi_{-c}=(-\eta)^{1+d\over2}\sum_s C_4^s({\bf k})H_{\nu_+}^2(-k\eta)a_+^s({\bf k})\equiv \sum_s \psi_{-c}^s\label{psi-cbd}
	\end{align}
	and, $C_3^s({\bf k})^*=0$ in \eq{psi+cd} and in \eq{psi-cd} resulting
	\begin{align}
		&\psi_{+c}^\dagger=i(-\eta)^{1+d\over2}\sum_s -a_+^s({\bf k})^\dagger C_4^s({\bf k})^*e^{i\pi\nu_+}{\gamma^i k_i\over k}H^1_{\nu_+}(-k\eta)\label{psi+cdb}\\
		&\psi_{-c}^\dagger=(-\eta)^{1+d\over2}\sum_s C_4^s({\bf k})^*H_{\nu_-}^1(-k\eta)a_+^s({\bf k})^\dagger \label{psi-cdb}
	\end{align}
	where we use the subscript $c$ to denote the classical value of the quantity. Now, we are left with two independent complex coefficients $C_4$ and $C_4^*$, for each $s$, which can be determined by specifying any of the following pairs at the boundary $\eta_B$ : $\{\psi_+,\psi_+^\dagger\} , \{\psi_-,\psi_-^\dagger\}$, $\{\psi_-,\psi_+^\dagger\}, \{\psi_+,\psi_-^\dagger\}$.
	 Note that the wavefunctional should be represented in terms of that pair of variables chosen to be specified on the boundary. Also note that each pair is related to the other by their saddle point value, see \eq{Mcl} for clarity.\footnote{The situation can be thought very similar to scalar theory where the field and its canonical momentum are related by classical saddle and the wavefunction of a state is represented in either basis.}
	  
	  Let us consider the case where we choose to specify $\{\psi_-,\psi_+^\dagger\}$ on the boundary. 
	  Now, varying the action \eq{frdiract} w.r.t. $\psi,{\bar\psi}$ respecting this boundary condition and using \eq{splitpsi}, \eq{barpsisplit} yields at saddle point
	\begin{equation}
		\label{delsb}
		\delta S_D|_c=  {1\over2}\int_{\Sigma} d^dx \sqrt{\gamma} n_\mu ie^\mu_a(\psi_{+c}^\dagger-\psi_{-c}^\dagger)\gamma^a\delta\psi_+ 
	\end{equation}
	where $\Sigma$ specifies an arbitrary time-slice $\eta$ as the boundary.
	Using \eq{nbeinds} and
	\begin{equation}
		n_\mu={R\over\eta}\delta_\mu^0,
	\end{equation}
	and the orthogonality between projection spaces, we get
	\begin{equation}
		\label{delsb+}
		\delta S_D|_c=  -{1\over2}\int_{\Sigma} d^dx \sqrt{\gamma}  i{\psi}_{+c}^\dagger\delta\psi_+
	\end{equation}
	Adding the following boundary term
	\begin{equation}
		\tilde{S}_B= {1\over2}\int_{\Sigma} d^dx \sqrt{\gamma}  i{\chi}_+^\dagger \psi_+ 
	\end{equation}
	to the Dirac action $S_D$ \eq{frdiract} where $\chi_\pm$ is introduced as an auxillary Grassman valued spinor field, 
	we see that the total variation $\delta(S_D+\tilde{S}_B)$ vanishes by identifying 
	\begin{equation}
		\label{chitopsi}
		\chi_-=\psi_{- c} ~~,~~ {\chi}_+^\dagger={\psi}_{+ c}^\dagger	
	\end{equation}
	Hence it is legitimate to consider $S[\psi,{\bar\psi}]$ in the Path integral \eq{pifer} to be the total action $\tilde S = S_D+\tilde{S}_B$.\\
	At saddle point approximation, the path integral \eq{pifer} can be written as
	\begin{equation}
		\Psi=e^{i(S_D+\tilde{S}_B)}
	\end{equation}
	where $\Psi$ is the wave functional of Bunch Davies vacuum and at tree-level, $S_D+\tilde{S}_B$ carries only the on-shell contribution determined by the classical values of $\psi$, i.e. $\psi_c$. Therefore, the tree-level result of the above equation is given as
	\begin{equation}
		\label{pic1}
		\Psi=\exp\brt{-\int_{\Sigma} d^dx {R^d\over 2(-\eta)^d} {\chi}_{+}^\dagger \psi_{+c} }
	\end{equation} 
	Noting that $\psi_{+c}$ can be related to $\psi_{-c}$ from \eq{psi+cbd} and \eqref{psi-cbd} as
	\begin{equation}
		\label{Mcl}
		\psi_{-c}^s({\bf k},\eta)={\mathbb M}({\bf k},\eta)\psi_{+c}^s({\bf k},\eta)~~~~~~{\mathbb M}({\bf k},\eta)\equiv ie^{i\pi\nu_-}{H^2_{\nu_+}(-k\eta)\over H^2_{\nu_-}(-k\eta)}(-{\gamma^ik_i\over k})
	\end{equation}
	one get the wave functional using \eq{Mcl} in \eq{pic1} as
	\begin{equation}
		\label{psibypsich+-}
		\Psi[{\chi_+^\dagger},\chi_-]=\exp\brt{-{R^d\over2}\sum_{s_1s_2}\int {d^dk\over(2\pi)^d} (-{\eta})^{-d}{\chi}_{+}^{s_1\dagger}({\bf k},{\eta}) {{\mathbb M}^{-1}({\bf k},{\eta})}\chi_{-}^{s_2}({\bf k},{\eta})}
	\end{equation}
	Similarly, specifying $\{\psi_+,\psi_-^\dagger\}$ on the boundary, and adding the following boundary term,
	\begin{equation}
		\tilde{S}_B= {1\over2}\int_{\Sigma} d^dx \sqrt{\gamma}  i{\chi}_-^\dagger \psi_- 
	\end{equation}
	one can arrive at the following representation of the wavefunctional
	\begin{equation}
		\label{psibypsich-+}
		\Psi[{\chi_-^\dagger},\chi_+,\eta]=\exp\brt{-{R^d\over2}\sum_{s_1s_2}\int {d^dk\over(2\pi)^d} (-{\eta})^{-d}{\chi}_{-}^{s_1\dagger}({\bf k},{\eta}) {{\mathbb M}({\bf k},{\eta})}\chi_{+}^{s_2}({\bf k},{\eta})}
	\end{equation}	 
	 The other two representations are found by specifying $\{\psi_+,\psi_+^\dagger\}$ and $\{\psi_-,\psi_-^\dagger\}$ on the boundary with addition of appropriate boundary actions,
	 \begin{align}
	 		&\tilde{S}_B= {1\over2}\int_{\Sigma} d^dxd^dy \sqrt{\gamma}  i{\chi}_+^\dagger({\bf y}){\tilde{\mathbb M}}^\dagger ({\bf y}-{\bf x}) \psi_-({\bf x})+{1\over2}\int_{\Sigma} d^dx \sqrt{\gamma}  i{\chi}_+^\dagger({\bf x}) \chi_+({\bf x})\\
	 		&\tilde{S}_B= {1\over2}\int_{\Sigma} d^dxd^dy \sqrt{\gamma}  i{\chi}_-^\dagger({\bf y}){(\tilde{\mathbb M}^{-1}({\bf y}-{\bf x}))}^\dagger \psi_+({\bf x}) +{1\over2}\int_{\Sigma} d^dx \sqrt{\gamma}  i{\chi}_-^\dagger({\bf x}) \chi_-({\bf x}) 
	 \end{align}
	 respectively and given by
	 \begin{align}
	 	&\Psi[\chi_+^\dagger,\chi_+,\eta]=\exp\brt{-{R^d\over2}\sum_{s_1s_2} \int{d^dk\over (2\pi)^d} (-\eta)^{-d} \chi_+^{s_1\dagger}({\bf k},\eta)[{\mathbb I}+{\mathbb M}^\dagger({\bf k},{\eta}) {\mathbb M}({\bf k},{\eta})]\chi_+^{s_2}({\bf k},\eta)} \label{psibypsich++}\\
	 	&\Psi[\chi_-^\dagger,\chi_-,\eta]=\exp\brt{-{R^d\over2}\sum_{s_1s_2} \int{d^dk\over (2\pi)^d} (-\eta)^{-d} \chi_-^{s_1\dagger}({\bf k},\eta)[{\mathbb I}+({\mathbb M}({\bf k},{\eta}){\mathbb M}^\dagger({\bf k},{\eta}))^{-1}] \chi_-^{s_2}({\bf k},\eta)} \label{psibypsich--}
	 \end{align}
 	where
 	\begin{equation}
 		{\tilde{\mathbb M}}({\bf x}-{\bf y})=\int {d^dk\over (2\pi)^d} {\mathbb M} ({\bf k}) e^{i{\bf k}.({\bf x}-{\bf y})}
 	\end{equation} 
	\section{Aspects of Holography}
	\label{sec.holo}
	The basic dictionary \cite{NG_malda} for the holography in dS space for the hologram living at ${\cal I}^+$ is that the wave function of a state in the bulk is equal to the partition function of the boundary field theory under appropriate identification of sources with the boundary values of the fields.
	Aspects of this version of holography has been studied in \cite{roy_dscft} for scalar field in dS$_{d+1}$ pertaining to both complementary and principal series representations. 
	In this section, we extend the study for Fermionic Fields.
	
	It is assumed that Wave functional of a state in Fermionic field theory can be expressed schematically as sum of terms consisting various powers of sources $\rho_\pm ({ \bf x})$ \footnote{$\rho_{\pm}$ is introduced as a general notation for spinors in \eq{Psigen}--\eq{dercoind} and $\langle\zeta_+\zeta_-^\dagger\cdots\rangle$ denotes a general matrix-valued quantity appearing in the coefficient.} as
	\begin{align}
		\log{\Psi}&=\int d^dx_1d^dx_2 \rho^{\dagger}_+({\bf x}_1)\langle\zeta_{+}({\bf x}_1)\zeta_{-}^\dagger({\bf x}_2)\rangle \rho_-({\bf x}_2) \nonumber\\&+ {1\over 2!}\int d^dx_1d^dx_2d^dx_3d^dx_4\rho_+^{\dagger}({\bf x}_1)\rho^{\dagger}_+({\bf x}_3)\langle\zeta_{+}({\bf x}_1)\zeta_{-}^\dagger({\bf x}_2)\zeta_{+}({\bf x}_3)\zeta_{-}^\dagger({\bf x}_4)\rangle \rho_-({\bf x}_2)\rho_-({\bf x}_4) + \cdots \label{Psigen} 
	\end{align}
	The statement of holography in this case is that the coefficient functions given from \eq{Psigen} as
	\begin{equation}
		\label{derco}
		{\delta^n\log[\Psi] \over \delta \rho_+^{\dagger}({\bf x}_1)\delta\rho_-({\bf x}_2)\delta\rho_+^{\dagger}({\bf x}_3)\delta\rho_-({\bf x}_4)\cdots} =(-1)^{n/2}\langle \zeta_{+}({\bf x}_1)\zeta_{-}^\dagger({\bf x}_2)\zeta_{+}({\bf x}_3)\zeta_{-}^\dagger({\bf x}_4) \cdots\rangle
	\end{equation}
	are in fact the correlation functions of various fields in the dual CFT. The phase factor $(-1)^{n/2}$ appears in the above equation due to the functional differentiation w.r.t. $\rho_{-}$ which is a Grassmann variable.\footnote{Note that, since $n$ being always an even number, the phase contributes up to an overall sign in the resulting correlation function.} To be more precise about the terms in \eq{Psigen} and \eq{derco}, we write the terms in \eq{Psigen} breaking into their components as
	\begin{align}
		&\rho^\dagger_+({\bf x}_1)\langle\zeta_+({\bf x}_1)\zeta_-^\dagger({\bf x}_2)\rangle \rho_-({\bf x}_2) \leftrightarrow \rho^{\dot{\mu}}_+({\bf x}_1)^*\langle\zeta_+({\bf x}_1)\zeta_-^\dagger({\bf x}_2)\rangle_{\dot{\mu}}^{\dot{\nu}} \rho_{-\dot{\nu}}({\bf x}_2)\\
		&\rho_+^\dagger({\bf x}_1)\rho^\dagger_+({\bf x}_3)\langle\zeta_+({\bf x}_1)\zeta_-^\dagger({\bf x}_2)\zeta_+({\bf x}_3)\zeta_-^\dagger({\bf x}_4)\rangle \rho_-({\bf x}_2)\rho_-({\bf x}_4) \nonumber\\& \leftrightarrow \rho_+^{\dot\mu}({\bf x}_1)^*\rho^{\dot\nu}_+({\bf x}_3)^*\langle\zeta_+({\bf x}_1)\zeta_-^\dagger({\bf x}_2)\zeta_+({\bf x}_3)\zeta_-^\dagger({\bf x}_4)\rangle_{{\dot\mu}{\dot\nu}}^{{\dot\alpha}{\dot\beta}} \rho_{-{\dot\alpha}}({\bf x}_2)\rho_{-{\dot\beta}}({\bf x}_4) 
	\end{align}
	Similarly, \eq{derco} with explicit index structure is given by
	\begin{equation}
		\label{dercoind}
		{\delta^n\log[\Psi] \over \delta \rho_{+}^{\dot \mu}({\bf x}_1)^*\delta\rho_{-{\dot \nu}}({\bf x}_2)\delta\rho_{+}^{\dot \alpha}({\bf x}_3)^*\delta\rho_{-{\dot \beta}}({\bf x}_4)\cdots} = (-1)^{n/2}\langle \zeta_{+}({\bf x}_1)\zeta_{-}^\dagger({\bf x}_2)\zeta_{+}({\bf x}_3)\zeta_{-}^\dagger({\bf x}_4) \cdots\rangle_{{\dot\mu}{\dot \alpha}\cdots}^{{\dot \nu}{\dot \beta}\cdots}
	\end{equation}
	Now, we show, for free theory, how the boundary values of bulk Fermionic field can be related to sources in the dual CFT. We begin with the wavefunction given by \eq{psibypsich+-},
	\begin{equation}
		\label{psibypsi}
		\Psi[{\psi_+^\dagger},\psi_-]=\exp\brt{-{R^d\over2}\sum_{s_1s_2}\int {d^dk\over(2\pi)^d} (-{\eta})^{-d}{\psi}_{+c}^{s_1\dagger}({\bf k},{\eta}) {{\mathbb M}^{-1}({\bf k},{\eta})}\psi_{-c}^{s_2}({\bf k},{\eta})}
	\end{equation}
	The late-time limit of \eq{psibypsi} is given by
	\begin{equation}
		\label{latewf}
		\Psi=\exp\brt{{R^d\over 2^{1+2iMR}}\sum_{s_1s_2} \int {d^dk\over (2\pi)^d} (-\eta_B)^{-d+2iMR} \psi_{+c}^{s_1\dagger} ({\bf k},\eta_B) \brf{e^{-\pi MR}{\Gamma[\nu_-]\over \Gamma[\nu_+]}{\gamma^ik_i\over k}k^{2iMR}}\psi^{s_2}_{-c}({\bf k},\eta_B)}
	\end{equation}
	where we use $\eta\rightarrow0$ limit of ${\mathbb M}$ defined in \eq{Mcl} and $\eta=\eta_B$ as late-time cut-off. Now, in terms of the quantities
	\begin{equation}
		\label{rhopm_s}
		\rho^s_+({\bf k})\equiv (-\eta_B)^{-{d\over2}-iMR}\psi^s_{+c}({\bf k},\eta_B)~~,~~\rho^s_-({\bf k})\equiv(-\eta_B)^{-{d\over2}+iMR}\psi^s_{-c}({\bf k},\eta_B)
	\end{equation}
	we can write \eq{latewf} as
	\begin{equation}
		\label{latewfrho}
		\Psi[\rho_+^\dagger,\rho_-]=\exp\brt{{R^d\over 2^{1+2iMR}}\sum_{s_1s_2}\int {d^dk\over (2\pi)^d}\rho_{+}^{s_1\dagger} ({\bf k}) \brf{e^{-\pi MR}{\Gamma[\nu_-]\over \Gamma[\nu_+]}{\gamma^ik_i\over k}k^{2iMR}}\rho^{s_2}_{-}({\bf k})}
	\end{equation}
	Also, in position space
	\begin{equation}
		\label{latewfrhox}
		\Psi[\rho_+^\dagger,\rho_-]=\exp\brt{{R^d\over2}\sum_{s_1s_2}\int {d^dx_1}{d^dx_2}\rho_{+}^{s_1\dagger} ({\bf x}_1) \brf{{i\over \pi^{d/2}}e^{-\pi MR}{\Gamma[{d\over2}+\nu_+]\over\Gamma[\nu_+]}}{\gamma^i(x_1^i-x_2^i)\over|{\bf x}_1-{\bf x}_2|^{d+2iMR+1}}\rho^{s_2}_{-}({\bf x}_2)}
	\end{equation}
	
	To derive the coefficient function from \eq{latewfrhox} by varying the wavefunctional, see \eq{derco}, w.r.t. the sources $\rho_\pm$ identified in \eq{rhopm_s}, we first have to count the number of independent components in $\rho_\pm$. It is because in the course of variation w.r.t. a component of $\rho_\pm$, other components must not vary. Note that number of dimensions $d_{\gamma}$ of irreducible representations (IrReps) of $\gamma$-matrices are given by the no. of  dimensions $n$ of the underlying manifold as  
	\begin{equation}
		\label{dgam}
		d_{\gamma}=2^{\brt{n\over 2}}\times 2^{\brt{n\over 2}}
	\end{equation}
	where $f(x)=[x]$ is the floor function.
	The sources $\rho_+,\rho_-$ are identified with the projections of bulk field $\psi$ onto the eigenspaces of $\gamma^0$. Therefore each of $\rho_\pm$ and $\rho_\pm^\dagger$ has $2^{[{d+1\over2}]-1}$ no. of independent components. One can also think it in the other way -- for each $s$, there are two independent variables to vary -- $C_4^s$, $C_4^{s*}$, see section \ref{sec.bdwf}. Therefore by \eq{dercoind}, we see that the resulting matrix has rank half of that of ${\gamma^ik^i\over k}$ in the bulk. 
	
	It is convenient to work in Dirac representation, see Appendix \ref{App.subtled}, for the purpose of variation w.r.t. independent variables following \eq{dercoind}. In this representation, one can write by \eq{psipmaseig} and \eq{rhopm_s}, 
	\begin{equation}
		\label{rhohat}
		\rho_-=\left[
		\begin{array}{c}
			0\\
			{\hat\rho}_{-}\\
		\end{array}
		\right]~~,~~
		\rho_+=\left[
		\begin{array}{c}
			{\hat\rho}_{+}\\
			0\\
		\end{array}
		\right]
	\end{equation}  
	where all the entries of $\hat\rho_{-}$ and $\hat\rho_{+}^\dagger$ are mutually independent. This reduces \eq{latewfrho} to,   
	\begin{equation}
		\Psi=\exp\brt{{R^d\over 2^{1+2iMR}}\int {d^dk\over (2\pi)^d}{\hat\rho}_{+}^{{\dot\alpha}} ({\bf k})^* \brf{e^{-\pi MR}{\Gamma[\nu_-]\over \Gamma[\nu_+]}{\gamma_{+-}^ik_i\over k}k^{2iMR}}_{\dot\alpha}^{\dot\beta}{\hat\rho}_{-{\dot\beta}}({\bf k})}
	\end{equation}
	where ${\gamma}^i_{+-}$ is the `top-right quarter block' of $\gamma^i$.
	This implies that the coefficient function in momentum space using \eq{dercoind} is to be in the form
	\begin{equation}
		\label{FFmom}
		\langle{\hat\zeta}_{+}({\bf k}){\hat\zeta}_{-}^\dagger({\bf k})\rangle = -{R^d\over2^{1+2iMR}}\brf{e^{-\pi MR}{\Gamma[\nu_-]\over \Gamma[\nu_+]}}{\gamma^i_{+-}k_i\over k}k^{2iMR}
	\end{equation}
	In position space, above result reduces to
	\begin{equation}
		\label{FFpos}
		\langle{\hat\zeta}_{+}({\bf x}_1){\hat\zeta}_{-}^\dagger({\bf x}_2)\rangle =- {R^d\over2}\brf{{i\over \pi^{d/2}}e^{-\pi MR}{\Gamma[{d\over2}+\nu_+]\over\Gamma[\nu_+]}}{{\gamma}^i_{+-}(x_1^i-x_2^i)\over|{\bf x}_1-{\bf x}_2|^{d+2iMR+1}}
	\end{equation}
	From \eq{FFpos}, we argue that the wavefunction coefficient behaves like the two point function of a Fermionic CFT operator where the dimension of the ${\hat\zeta}_+$ is given by
	\begin{equation}
		\label{Fdim}
		\Delta_+={d\over2}+iMR
	\end{equation}
	which is found to be complex. We refer appendix \ref{App.trandir}, particularly \eq{dilzeta} and \eq{sctbf}, where we show by deriving the transformation of ${\hat\zeta}_+$ under conformal isometries that $\Delta_+$ is indeed the conformal dimension of ${\hat\zeta}_+$. This complex value is also expected from the analytic continuation, $L\rightarrow iR$, see \eq{analcont}, from the conformal dimension of the boundary theory of AdS space of radius $L$, \eq{cdimads}, see also \cite{Henning_ferads, Muckviswa_ferads}, and thereby the putative CFT on ${\cal I}^+$ is to be non-unitary.\\
	It is also to note that the dimension of ${\hat\zeta}_-$ is then given by
	\begin{equation}
		\label{dim-}
		\Delta_-={d\over2}-iMR
	\end{equation}
	And, the analytic continuation $L\rightarrow -iR$, which is argued to be the correct continuation in the lower Poincare patch in \cite{roy_dscft}, reproduces \eq{dim-} from \eq{cdimads}.
	
	One does also encounter complex value of the conformal dimension $\Delta$ of the dual theory on ${\cal I}^+$ corresponding to Bunch-Davies vacuum for scalar fields of mass $M>{d\over2R}$ in de Sitter space  by identifying the boundary source as the eigenvalue of the coherent state basis \cite{roy_dscft}, also see \cite{Isono_dscft} for related discussion in terms of mixed boundary conditions. The identification of source, that would lead to the conformal dimension to be conjugate to $\Delta$, is ruled out based on the fact that the identification would lead to non-normalisable wave functional, leading to an unambiguous prescription of dS/CFT duality. However, for scalar fields of mass $M<{d\over 2R}$, the identification of source is more analogous to that in AdS/CFT. In appendix \ref{App.int}, we evaluate the late-time wave function coefficient including interaction of Fermionic field with one scalar field, generally known as Yukawa interaction, in the leading order of coupling constant. 
	We show, for both the cases, that under appropriate identification of sources, the coefficient \eq{3ptpos} takes the form of CFT correlator where the boundary Fermionic field ${\hat\zeta}_\pm$ acts as a primary field with the conformal dimension $\Delta_\pm$ given in \eq{Fdim} and \eq{dim-}.\\~\\
	In Fermionic theory,
	one can also think of considering the representation of the wave functional as given in \eq{psibypsich-+} restated below
	\begin{equation}
		\label{psibypsial}
		\Psi[{\psi_-^\dagger},\psi_+]=\exp\brt{-{ R^d\over2}\sum_{s_1s_2}\int {d^dk\over(2\pi)^d} (-{\eta})^{-d}{\psi}_{-c}^{s_1\dagger}({\bf k},\eta) {{\mathbb M}({\bf k},{\eta})}\psi_{+c}^{s_2}({\bf k},\eta)}
	\end{equation}
	Under the identification \eq{rhopm_s}, one can then obtain the late time wave functional in momentum space as
	\begin{equation}
		\Psi[\rho_+^\dagger,\rho_-]=\exp\brt{-{R^d\over 2^{1-2iMR}}\sum_{s_1s_2}\int {d^dk\over (2\pi)^d}\rho_{-}^{s_1\dagger} ({\bf k}) \brf{e^{\pi MR}{\Gamma[\nu_+]\over \Gamma[\nu_-]}{\gamma^ik_i\over k}k^{-2iMR}}\rho^{s_2}_{+}({\bf k})}
	\end{equation}
	and, in position space as
	\begin{equation}
		\label{latewfalrho}
		\Psi[\rho_-^\dagger,\rho_+]=\exp\brt{{R^d\over 2}\sum_{s_1s_2}\int {d^dx_1}{d^dx_2}\rho_{-}^{s_1\dagger} ({\bf x}_2) \brf{{i\over \pi^{d/2}}e^{\pi MR}{\Gamma[{d\over2}+\nu_-]\over\Gamma[\nu_-]}}{\gamma^i(x_1^i-x_2^i)\over|{\bf x}_1-{\bf x}_2|^{d-2iMR+1}}\rho^{s_2}_{+}({\bf x}_1)}
	\end{equation}
	Now, following the similar arguments given in case of $\Psi[\psi_{+}^\dagger,\psi_{-}]$, we get by \eq{dercoind},
	\begin{equation}
		\label{wfcal}
		\langle {\hat\zeta}_{-}({\bf x}_2){\hat\zeta}_{+}^\dagger({\bf x}_1)\rangle=-{R^d\over2}\brf{{i\over \pi^{d/2}}e^{\pi MR}{\Gamma_[{d\over2}+\nu_-]\over\Gamma[\nu_-]}}{\gamma_{-+}^i(x_1^i-x_2^i)\over|{\bf x}_1-{\bf x}_2|^{d-2iMR+1}}
	\end{equation}
	where $\gamma_{-+}^i$ is the `bottom-left quarter block' of $\gamma^i$.
	We can immediately see, perhaps by comparing with \eq{FFpos}, that the wavefunction coefficient behaves like CFT two point function with fields ${\hat\zeta}_+,{\hat\zeta}_-$ of conformal dimensions $\Delta_+,\Delta_-$, \eq{Fdim} and \eq{dim-}, respectively.

	We therefore suggest that the dual theory on ${\cal I}^+$ admits two independent boundary spinor fields ${\hat\zeta}_+$ and ${\hat\zeta}_-$ of conformal dimensions $\Delta_+,\Delta_-$ respectively, which are conjugate to each other, corresponding to a single Dirac field $\psi$ in the bulk. It is in contrast with EAdS/CFT as follows. In EAdS/CFT, the partition function in the bulk is suggested to be equal to the partition function of the boundary CFT \cite{Gubser:1998bc,Witten:1998qj}. The correlation function given by the bulk partition function in EAdS space takes the form \cite{Henneaux_ferads,Henning_ferads}
	\begin{equation}
		\langle \zeta({\bf x}) \zeta^\dagger({\bf y})\rangle \sim {\gamma^i(x^i-y^i)\over|{\bf x}-{\bf y}|^{d+2ML+1}}
	\end{equation}
	where $\zeta$ having the conformal dimension $\Delta={d\over2}+ML$, \eq{cdimads}, is assigned to be the boundary spinor field dual to the bulk Dirac field.

	It is also interesting to note that both representations \eq{psibypsi} and \eq{psibypsial} of the wave functional are normalisable, and produce the two point functions $\langle {\hat\zeta}_+({\bf x}){\hat\zeta}_-^\dagger({\bf y})\rangle$ and $\langle {\hat\zeta}_-({\bf x}){\hat\zeta}_+^\dagger({\bf y})\rangle$ respectively. One can also think of constructing other two possible two-point correlation functions  $\langle {\hat\zeta}_+({\bf x}){\hat\zeta}_+^\dagger({\bf y})\rangle$ and $\langle {\hat\zeta}_-({\bf x}){\hat\zeta}_-^\dagger({\bf y})\rangle$ in the boundary theory. From the bulk wavefunctional, using \eq{rhopm_s} at late time, these correlations can be found in the wavefunction coefficient of $\Psi[\psi_+^\dagger,\psi_+]$ and $\Psi[\psi_-^\dagger,\psi_-]$ respectively as
	\begin{align}
		&\langle {\hat\zeta}_{+}({\bf x}_1){\hat\zeta}_{+}^\dagger({\bf x}_2)\rangle
		={R^d\over 2}(1+e^{2\pi MR})\delta({\bf x}_1-{\bf x}_2){\mathbb I}\\
		&\langle {\hat\zeta}_{-}({\bf x}_1){\hat\zeta}_{-}^\dagger({\bf x}_2)\rangle
		={R^d\over2}(1+e^{-2\pi MR})\delta({\bf x}_1-{\bf x}_2){\mathbb I}
	\end{align} 
	where $\mathbb I$ is the identity matrix of appropriate dimension. These correlations are found to be local in position space. Therefore these two correlators alone are not sufficient to determine the conformal dimension of the boundary fields ${\hat\zeta}_+,{\hat\zeta}_-$.

	
	
	We conclude the section commenting on the ${\hat\zeta}_+,{\hat\zeta}_-$ fields on the boundary in $d={\rm odd}$ and $d={\rm even}$. In the above discussions, we see that the correlators obtained from the wavefunctionals \eq{psibypsi} and \eq{psibypsial} involve off-diagonal blocks of Dirac $\gamma^i$ matrices defined in the bulk. In particular, $\langle{\hat\zeta}_+({\bf x}){\hat\zeta}_-^\dagger({\bf y})\rangle$ involves the top-right block and $\langle{\hat\zeta}_-({\bf x}){\hat\zeta}_+^\dagger({\bf y})\rangle$ involves the bottom-left block. In $d={\rm odd}$, these off-diagonal blocks are proportional to Dirac matrices defined on the boundary Euclidean manifold. Therefore, it can be argued that ${\hat\zeta}_+, {\hat\zeta}_-$ behave as two independent primary Dirac fields defined in the boundary theory. In $d={\rm even}$, dimension of Dirac-$\gamma$ on the boundary is the same as that in the bulk. Therefore the fields ${\hat\zeta}_\pm$ in this case behave as independent left and right handed primary Weyl spinors with different conformal dimensions. Appendix \ref{App.subtled} includes representations of $\gamma$ matrices in $d=3,4$ to bring more clarity in this discussion.

	\section{Ward Identities}
	\label{Sec.WI}
	Having identified the source of the dual field theory in terms of the asymptotic value of the bulk field, it is now important to discuss Ward identities that the resulting correlation functions should satisfy, see \cite{Maldacena:2011nz, Ghosh:2014kba, Kundu:2015xta, Bzowski:2023nef} in the context of scalar fields. As discussed in \cite{roy_dscft}, also see \cite{Freedman:1998tz}, the issue of meeting Ward identities in holography crucially depends on the nature of the source identification. In case of scalar fields, for example, non-local identification of source in position space is found to spoil the Ward identities \cite{roy_dscft}. However, in case of Fermionic field, identification of the source in terms of the bulk field is local in position space, \eq{rhopm_s}, which makes the fact of Ward identities holding true for correlation functions transparent as we will explain in this section.
	
	Ward identities are statements about symmetries that are present in the theory. The identities result from the requirement of invariance of the wave functional under the action of underlying symmetry transformations. More explicitly, working in ADM gauge, the metric can be written including perturbation as
	\begin{equation}
		ds^2=-{d\eta^2\over \eta^2}+{1\over \eta^2}{\hat g}_{ij}dx^idx^j
	\end{equation}
	Note that in the above metric, we set lapse function $N^2={1\over \eta^2}$ and shift function $N^i=0$. Equations of motion obtained varying the action w.r.t. these $N$ and $N^i$  yield the Hamilton and momentum constraints which are symmetry constraints leading to the conditions of spatial and time reparametrisation. States satisfying these conditions are the solutions of WdW equation which includes Bunch Davies vacuum as a solution. This ensures that Bunch Davies vacuum is to be invariant under residual gauge transformations preserving ADM gauge. How the invariance of the wave functional can be implemented in its path integral formulation in general, and more importantly, at the saddle point approximation has been discussed in \cite{roy_dscft}.
	
	Now considering a physical state in Fermionic field theory represented in terms of projection of bulk field $\psi$ and, more generally, metric components ${\hat g}_{ij}$, let us take the value of the wavefunctional at late time cut-off $\eta=\eta_B$ as $\Psi[\psi_+^\dagger, \psi_-, {\hat g}_{ij},\eta_B]$.  Under general coordinate transformation which assumes
	\begin{equation}
		\psi \rightarrow\psi+\delta\psi~~;~~{\hat g}_{ij}\rightarrow {\hat g}_{ij}+\delta {\hat g}_{ij}~~;~~\eta_B\rightarrow\eta_B+\delta \eta_B
	\end{equation}
	invariance of the wave functional states
	\begin{equation}
		\label{invcondpsi}
		\Psi[\psi_+^\dagger +\delta\psi_+^\dagger,\psi_- +\delta\psi_-,{\hat g}_{ij}+\delta {\hat g}_{ij},\eta_B+\delta \eta_B]=\Psi[\psi_+^\dagger,\psi_-,{\hat g}_{ij},\eta_B]
	\end{equation}
	Transformation of metric under coordinate transformation has been derived in Appendix D of \cite{roy_dscft}. 
	However, it is worth to introduce here transformation of Fermionic field. 
	Under general infinitesimal coordinate transformation given by
	\begin{equation}
		\label{genco}
		x'^\mu=x^\mu+\xi^\mu(x)
	\end{equation}
	the transformation of Fermionic field is given in \cite{Hinterbichler:2026xqf} as \footnote{
		We make two comments at this point. First, the signature of the commutator term in \eq{gendelwf} is taken opposite to the cited reference due to the opposite sign convention in the anti-commutation of $\gamma$-matrices, \eq{gacomdag}. Second, in the given reference, the transformation of Fermionic field is given considering $\xi^\mu$ as killing vector under the dS isometry group. We extend the result defining the transformation rule for Fermionic field under generic diffeomorphism.  
	}
	\begin{equation}
		\label{gendelwf}
		\delta\psi(x)\equiv\psi'(x)-\psi(x)=-\xi^\mu\nabla_\mu\psi(x)+{1\over8}\nabla_\mu\xi_\nu e^\mu_a e^\nu_b[\gamma^a,\gamma^b]\psi(x)
	\end{equation}

	Now, given the relation between source and the bulk field, one can evaluate corresponding transformation of the source. Finally obtaining the invariance condition \eq{invcondpsi} in terms of sources, one can argue that the wavefunction coefficients arising from the expansion of wave functional in terms of sources satisfy CFT Ward identities.
	
	To note, we have suppressed the index for spin degenaracies to save unnecessary clutter in the following discussions in this section.
	
	\subsection{Identity for Local Lorentz Transformation}
	As we discussed in section \ref{sec.clsol} and Appendix \ref{App.trandir} in more details, formalism of Fermionic field theory in curved space-time explicitly requires the notion of local orthonormal frame at each point on the manifold. The frame can be defined through the quantity called `n-bein' which is uniquely defined up to a local Lorentz transformation. One can therefore think about performing only a local Lorentz transformation and ask for the invariance of the wavefunctional.
	
	Parameterising an infinitesimal local Lorentz transformation (ILLT) by an infintesimal but arbitrary quantity $\epsilon_{ab}(x)$, the Dirac field is found to be transformed under the ILLT as,
	\begin{equation}
		\label{deldirillt}
		\delta\psi(x)=-{1\over8}\epsilon_{ab}({\bf x},\eta)[\gamma^a,\gamma^b]\psi(x)
	\end{equation}
	More discussions are available in Appendix \ref{App.trandir}. Using \eq{splitpsi}, we find $\psi_\pm$ transforming as
	\begin{equation}
		\delta\psi_+(x)=P_+\delta\psi(x)=-{1\over4}\epsilon_{0i}({\bf x},\eta)[\gamma^0,\gamma^i]\psi_-(x) - {1\over8}\epsilon_{ij}({\bf x},\eta)[\gamma^i,\gamma^j]\psi_+(x)
	\end{equation} 
	\begin{equation}
		\delta\psi_-(x)=P_-\delta\psi(x)=-{1\over4}\epsilon_{0i}({\bf x},\eta)[\gamma^0,\gamma^i]\psi_+(x) - {1\over8}\epsilon_{ij}({\bf x},\eta)[\gamma^i,\gamma^j]\psi_-(x)
	\end{equation} 
	It is interesting to observe in the above expression that if we fix $\epsilon_{0i}=0$ keeping $\epsilon_{ij}$ arbitrary,
	the transformation of $\psi_+$ (or, $\psi_-$) becomes independent of $\psi_-$ (or, $\psi_+$). 
	The source, defined in \eq{rhopm_s}, then also transforms as
	\begin{equation}
		\label{delrhoillt}
		\delta\rho_\pm({\bf x})=-{1\over8}\epsilon_{ij}({\bf x},\eta)[\gamma^i,\gamma^j]\rho_\pm({\bf x})
	\end{equation}
	Invariance of the wave functional \eq{Psigen} can therefore be stated as
	\begin{equation}
		\label{wfvarillt}
		\Psi[\rho_+^\dagger + \delta\rho_+^\dagger, \rho_- +\delta\rho_-, \eta_B]=\Psi[\rho_+^\dagger,\rho_-,\eta_B]
	\end{equation}
	
	Considering the wavefunctional in the form
	\begin{equation}
		{\Psi}[\rho_+^\dagger,\rho_-,\eta]=\exp\brt{\int d^dx_1d^dx_2 \rho^{\dagger}_+({\bf x}_1)\langle\zeta_{+}({\bf x}_1)\zeta_{-}^\dagger({\bf x}_2)\rangle \rho_-({\bf x}_2)}
	\end{equation}
	where $\rho_\pm$ assumes the Dirac structure \eq{rhohat}, the \eq{wfvarillt} leads to the identity involving two point function in the following integrated version 
	\begin{equation}
		\label{idllt}
		\int_x {\hat\rho}_+^\dagger({\bf x}_1) \epsilon_{ij}({\bf x}_1)[\gamma^i,\gamma^j]_{++}\langle \zeta_+({\bf x}_1)\zeta_-^\dagger({\bf x}_2) \rangle_{+-} {\hat\rho}_-({\bf x}_2)= \int_x {\hat\rho}_+^\dagger({\bf x}_1)\epsilon_{ij}({\bf x}_2)\langle \zeta_+({\bf x}_1)\zeta_-^\dagger({\bf x}_2) \rangle_{+-}[\gamma^i,\gamma^j]_{--}{\hat\rho}_-({\bf x}_2)
	\end{equation}
	where the subscripts `$++$', `$--$', `$+-$' denote the top-left, bottom-right, and top-right quarter block respectively of the associated matrix form. Following the discussions in section \ref{sec.holo}, the block $\langle\zeta_+({\bf x})\zeta_-^\dagger({\bf y})\rangle_{+-}$ is indeed the two point correlators $ \langle{\hat\zeta}_+({\bf x}){\hat\zeta}_-^\dagger({\bf y})\rangle$ of the boundary fields.

	\subsection{Ward Identity for Spatial Reparametrisation}
	Spatial reparametrisation is given by the infinitesimal coordinate change as
	\begin{equation}
		\label{sprep}
		x'^i=x^i+v^i({\bf x})~~,~~\eta'=\eta
	\end{equation}
	Comparing with \eq{genco}, $\xi^\mu(x)$ is given in this reparametrisation as,
	\begin{equation}
		\label{spdif}
		\xi^0({\bf x},\eta)=0~~,~~\xi^i({\bf x},\eta)=v^i({\bf x})
	\end{equation}
	Transformation of metric source under this reparametrisation is discussed in \cite{roy_dscft}. Here we will be brief on this. We consider the metric ${\hat g}_{ij}$ is expanded about its flat space value
	\begin{equation}
		{\hat g}_{ij}=\delta_{ij}+\gamma_{ij}
	\end{equation}
	where $\gamma_{ij}$ acts the source that generates stress-tensor components as response. Under diffeomorphism, the source ${\gamma_{ij}}$ transforms as
	\begin{equation}
		\label{delmetsp}
		\delta{\hat g}_{ij}=\delta\gamma_{ij}=-\del_iv_j({\bf x})-\del_j v_i({\bf x}) +{\rm homogeneous ~ contributions}
	\end{equation}
	where the homogeneous contributions are produced by terms proportional to ${\gamma_{ij}}$ in which we are not interested in. The transformation of metric determinant is given by
	\begin{equation}
		\label{delgsp}
		\delta\brf{ \sqrt{{\hat g}({\bf x})} }=-\del_iv^i({\bf x})
	\end{equation}
	The change in the Fermionic field given in \eq{gendelwf} can therefore be reduced, in this case, to \footnote{The indices $\{i,j\}$ are raised and lowered w.r.t. flat space metric $\delta_{ij}$.}
	\begin{equation}
		\label{delwfsp}
		\delta \psi({\bf x},\eta)=\brt{-v^i\del_i+{1\over 16}(\del_i v_j -\del_j v_i)[\gamma^i,\gamma^j] }\psi({\bf x},\eta)
	\end{equation}
	where we use \eq{cristds} and \eqref{spinconds}.
	Using \eq{delwfsp} and \eq{rhopm_s}, we get
	\begin{equation}
		\label{delrhosp}
		\delta \rho_\pm({\bf x})=\left({1\over16}(\del_i v_j-\del_j v_i)[\gamma^i,\gamma^j]-v^i({\bf x})\del_i\right)\rho_\pm({\bf x})
	\end{equation}
	Expansion of the wave functional in terms of sources \eq{Psigen} does also include the contribution from metric perturbation as
	\begin{align}
		\log{\Psi}&[\rho_+^\dagger,\rho_-,\gamma_{ij},\eta_B]=\int d^dx_1d^dx_2 \sqrt{\hat g({\bf x}_1)}\sqrt{\hat g({\bf x}_2)} \rho^\dagger_+({\bf x}_1)\langle\zeta_+({\bf x}_1)\zeta_-^\dagger({\bf x}_2)\rangle \rho_-({\bf x}_2) \nonumber\\&+ {1\over 2}\int d^dx_1d^dx_2d^dx_3 \sqrt{\hat g({\bf x}_1)}\sqrt{\hat g({\bf x}_2)}\sqrt{\hat g({\bf x}_3)}\gamma_{ij}({\bf x}_3)\rho_+^\dagger({\bf x}_1)\langle T^{ij}({\bf x}_3)\zeta_+({\bf x}_1)\zeta_-^\dagger({\bf x}_2)\rangle \rho_-({\bf x}_2) + \cdots \label{Psigenmet} 
	\end{align}
	Now from the invariance of the wave functional in terms of source
	\begin{equation}
		\Psi[\rho_+^\dagger+\delta\rho_+^\dagger,\rho_-+\delta\rho_-,\gamma_{ij}+\delta\gamma_{ij},\eta_B]=\Psi[\rho_+^\dagger,\rho_-, \gamma_{ij},\eta_B]
	\end{equation}
	we get by using \eq{delrhosp} and \eq{delgsp},
	\begin{equation}
		\label{addI}
		I_1+I_2+I_3=0
	\end{equation}
	where
	\begin{align}
		I_1=&-\int d^dx_1d^dx_2d^dx_3 (\del_i^{{\bf x}_3}v_j({\bf x}_3))\rho_+^\dagger({\bf x}_1)\langle T^{ij}({\bf x}_3)\zeta_+({\bf x}_1)\zeta_-^\dagger({\bf x}_2)\rangle\rho_-({\bf x}_2)\\
		I_2=&-\int d^dx_1d^dx_2[(\del_iv^i({\bf x}_1))\rho_+^\dagger({\bf x}_1)\langle \zeta_+({\bf x}_1)\zeta_-^\dagger({\bf x}_2)\rangle\rho_-({\bf x}_2)+(\del_iv^i({\bf x}_2))\rho_+^\dagger({\bf x}_1)\langle \zeta_+({\bf x}_1)\zeta_-^\dagger({\bf x}_2)\rangle\rho_-({\bf x}_2)]\nonumber\\
		&-\int d^dx_1d^dx_2[(v^i({\bf x}_1))\del_i\rho_+^\dagger({\bf x}_1)\langle \zeta_+({\bf x}_1)\zeta_-^\dagger({\bf x}_2)\rangle\rho_-({\bf x}_2)+\rho_+^\dagger({\bf x}_1)\langle \zeta_+({\bf x}_1)\zeta_-^\dagger({\bf x}_2)\rangle(v^i({\bf x}_2))\del_i\rho_-({\bf x}_2)]\\
		I_3=&-{1\over 8}\int d^dx_1d^dx_2\rho_+^\dagger({\bf x}_1)(\del_iv_j({\bf x_1}))[\gamma^i,\gamma^j]\langle \zeta_+({\bf x}_1)\zeta_-^\dagger({\bf x}_2)\rangle\rho_-({\bf x}_2)\nonumber\\
		&+{1\over 8}\int d^dx_1d^dx_2\rho_+^\dagger({\bf x}_1)(\del_iv_j({\bf x_2}))\langle \zeta_+({\bf x}_1)\zeta_-^\dagger({\bf x}_2)\rangle[\gamma^i,\gamma^j]\rho_-({\bf x}_2)
	\end{align}
	Let us systematically go through the integrals. Performing integration by parts and neglecting boundary terms, we can reduce $I_1$ and $I_2$ as
	\begin{align}
		I_1=&\int d^dx_1d^dx_2d^dx_3 v_j({\bf x}_3)\rho_+^\dagger({\bf x}_1)\del_i^{{\bf x}_3}\langle T^{ij}({\bf x}_3)\zeta_+({\bf x}_1)\zeta_-^\dagger({\bf x}_2)\rangle\rho_-({\bf x}_2)\\
		I_2=&\int d^dx_1d^dx_2[v^i({\bf x}_1)\rho_+^\dagger({\bf x}_1)\del_i^{{\bf x}_1}\langle \zeta_+({\bf x}_1)\zeta_-^\dagger({\bf x}_2)\rangle\rho_-({\bf x}_2)+v^i({\bf x}_2)\rho_+^\dagger({\bf x}_1)\del_i^{{\bf x}_2}\langle \zeta_+({\bf x}_1)\zeta_-^\dagger({\bf x}_2)\rangle\rho_-({\bf x}_2)]
	\end{align}
	$I_3$, on the other hand, can be reduced introducing $\delta$-function to
	\begin{align}
		I_3=&+{1\over 8}\int d^dx_1d^dx_2d^dx_3 v_j({\bf x_3}) \rho_+^\dagger({\bf x}_1)(\del_i^{{\bf x}_3}\delta({\bf x}_1-{\bf x}_3))[\gamma^i,\gamma^j]\langle \zeta_+({\bf x}_1)\zeta_-^\dagger({\bf x}_2)\rangle\rho_-({\bf x}_2)\nonumber\\
		&-{1\over 8}\int d^dx_1d^dx_2d^dx_3v_j({\bf x_3}) \rho_+^\dagger({\bf x}_1)(\del_i^{{\bf x}_3}\delta({\bf x}_2-{\bf x}_3))\langle \zeta_+({\bf x}_1)\zeta_-^\dagger({\bf x}_2)\rangle[\gamma^i,\gamma^j]\rho_-({\bf x}_2)
	\end{align}
	Performing an integration by parts and using \eq{idllt} with the consideration $\epsilon_{ij}=\del_iv_j-\del_jv_i$, we get
	\begin{align}
		I_3=&-{1\over 8}\int d^dx_1d^dx_2d^dx_3 v_j({\bf x_3}) \rho_+^\dagger({\bf x}_1)\delta({\bf x}_1-{\bf x}_3)[\gamma^i,\gamma^j]\del_i^{{\bf x}_1}\langle \zeta_+({\bf x}_1)\zeta_-^\dagger({\bf x}_2)\rangle\rho_-({\bf x}_2)\nonumber\\
		&+{1\over 8}\int d^dx_1d^dx_2d^dx_3v_j({\bf x_3}) \rho_+^\dagger({\bf x}_1)\delta({\bf x}_2-{\bf x}_3)\del_i^{{\bf x}_2}\langle \zeta_+({\bf x}_1)\zeta_-^\dagger({\bf x}_2)\rangle[\gamma^i,\gamma^j]\rho_-({\bf x}_2)
	\end{align}
	Hence by \eq{addI}, we get a relation between three point correlator involving stress-tensor component to two point correlators as
	\begin{align}
		\del_i^{{\bf x}_3}\langle T^{ij}({\bf x}_3){\hat\zeta}_+({\bf x}_1){\hat\zeta}_-^\dagger&({\bf x}_2)\rangle+\delta({\bf x}_1-{\bf x}_3)\brf{\delta^{ij}{\mathbb I}-{1\over 8}[\gamma^i,\gamma^j]}_{++}\del_i^{{\bf x}_1}\langle {\hat\zeta}_+({\bf x}_1){\hat\zeta}_-^\dagger({\bf x}_2)\rangle \nonumber\\&+\delta({\bf x}_2-{\bf x}_3)\del_i^{{\bf x}_2}\langle {\hat\zeta}_+({\bf x}_1){\hat\zeta}_-^\dagger({\bf x}_2)\rangle\brf{\delta^{ij}{\mathbb I}+{1\over 8}[\gamma^i,\gamma^j]}_{--}=0
	\end{align}
	which is the CFT Ward identity for translational invariance with operator ${\hat\zeta}_\pm$ transforming as primary spin-$1\over2$ fields. 
	
	\subsection{Ward Identity for Time Reparametrisation}
	Time reparametrisation is given by the infinitesimal coordinate transformation as
	\begin{equation}
		\label{trep}
		\eta'=\eta(1+\varepsilon({\bf x}))~~,~~{x'}^i=x^i+{1\over 2}\eta^2\del_i\varepsilon({\bf x})
	\end{equation}
	It is to note that time reparametrisation keeps the metric in ADM gauge asymptotically, i.e., $\eta_B\rightarrow0$. Therefore we can ignore, wherever appropriate, the change in spatial coordinates to the leading order in $\eta$. Comparing with \eq{genco}, we get
	\begin{equation}
		\xi^0({\bf x},\eta)=\eta\varepsilon({\bf x}) ~~,~~\xi^i({\bf x},\eta)=0+O(\eta^2)
	\end{equation}
	The change in the Fermionic field given in \eq{gendelwf} can therefore be reduced, in this case, to
	\begin{equation}
		\label{delwftime}
		\delta\psi({\bf x},\eta)=-\varepsilon({\bf x})\eta\del_\eta\psi({\bf x},\eta)-{\eta\over 8}\del_i\varepsilon({\bf x}) [\gamma^i,\gamma^0]\psi(x)
	\end{equation}
	Given the identification of source \eq{rhopm_s}, the corresponding transformation of $\rho_\pm$ can be found as
	\begin{equation}
		\label{rhotmrep}
		\delta\rho_\pm=-\varepsilon({\bf x})\Delta_\pm\rho_\pm +O(\eta)
	\end{equation}
	where $\Delta_\pm$ is given in \eq{Fdim} and \eqref{dim-}.

	The metric transforms under \eq{trep} as
	\begin{equation}
		\label{mettmrep}
		\delta{\hat g}_{ij}({\bf x})=2\varepsilon({\bf x})\delta_{ij}~~,~~\delta\sqrt{{\hat g}({\bf x})}=d\varepsilon({\bf x})\sqrt{{\hat g}({\bf x})}
	\end{equation}
	Invariance of action under time reparametrisation also requires the cut-off to be altered as 
	\begin{equation}
		\eta_B \rightarrow\eta_B(1+\varepsilon({\bf x}))
	\end{equation}
	This is discussed in more details in \cite{roy_dscft} for scalar fields and can be generalised to Fermionic fields also in the similar manner. The statement of invariance of the wave functional under these changes of fields and cut-off is given as
	\begin{equation}
		\Psi[\rho_+^\dagger+\delta\rho_+^\dagger,\rho_-+\delta\rho_-,\delta_{ij}+\delta{\hat g}_{ij},\eta_B+\delta \eta_B]=\Psi[\rho_+^\dagger,\rho_-,\delta_{ij},\eta_B]
	\end{equation}
	where $\Psi$ is given in \eq{Psigenmet}. Using \eq{mettmrep} and \eqref{rhotmrep}, we can derive from the above equation a relation between three point correlator involving trace of the stress tensor to two point correlator as
	\begin{equation}
		\langle T^i_i({\bf x}_3){\hat\zeta}_+({\bf x}_1){\hat\zeta}_-^\dagger({\bf x}_2) \rangle +\Delta_+\delta({\bf x}_1-{\bf x}_3)\langle {\hat\zeta}_+({\bf x}_3){\hat\zeta}_-^\dagger({\bf x}_2) \rangle +\Delta_+\delta({\bf x}_2-{\bf x}_3)\langle {\hat\zeta}_+({\bf x}_1){\hat\zeta}_-^\dagger({\bf x}_3) \rangle = 0
	\end{equation}
	This is indeed the Ward identity in CFT for local scale invariance.

	\subsection{Ward Identities for Conformal Symmetries}
	\label{sec.WIconf}
	Conformal symmetries are generated by a special class of diffeomorphisms under which the bulk metric remains invariant. The bulk with de Sitter geometry has isometry group $SO(d+1,1)$ and the transformations of coordinates that correspond to the isometries assume particular combinations of spatial and time reparametrisations. More precisely, $SO(d+1,1)$ is isomorphic to conformal group in $d$ dimensions whose generators give rise to the diffeomorphism of the form \eq{genco} with
	\begin{align}
		&{\rm Translation : }~ v^i({\bf x},\eta)=s^i~~~~~~~~~~~~~~~~~~~~~~~~~~~~~~~~~~~~~~~,~~\varepsilon({\bf x},\eta)=0\label{trdif}\\
		&{\rm Rotation~~~ : }~ v^i({\bf x},\eta)=s^i_{~j}x^j~~~~~~~~~~~~~~~~~~~~~~~~~~~~~~~~~~~~,~~\varepsilon({\bf x},\eta)=0\\
		&{\rm Dilatation~ : }~ v^i({\bf x},\eta)=sx^i~~~~~~~~~~~~~~~~~~~~~~~~~~~~~~~~~~~~~~,~~\varepsilon({\bf x},\eta)=s\\
		&{\rm SCT~~~~~~~~ : }~ v^i({\bf x},\eta)=2x^i(b_jx^j)-(x^jx_j-\eta^2)b^i~~~~~~~~~~,~~\varepsilon({\bf x},\eta)=2b^jx_j\label{sctdif}
	\end{align}
	where $s, s^i, s^i_j, b^i$ are infinitesimal quantities parameterising their respective diffeomorphisms.
	One can also verify that metric remains unchanged under the conformal transformations. Noting that at $\eta\rightarrow0$, $v^i$ and $\varepsilon$ do only depend on ${\bf x}$,  The source $\rho_\pm$ transforms as
	\begin{equation}
		\label{gendelrho}
		\delta \rho_\pm({\bf x})=\left({1\over16}(\del_i v_j-\del_j v_i)[\gamma^i,\gamma^j]-v^i({\bf x})\del_i-\varepsilon({\bf x})\Delta_\pm +O(\eta)\right)\rho_\pm({\bf x})
	\end{equation}
	where we put $v^i, \varepsilon$ appropriate to the transformations given in \eq{trdif} -- \eqref{sctdif}. Invariance of the wavefunction under the conformal symmetries
	\begin{equation}
		\Psi[\rho_+^\dagger+\delta\rho_+^\dagger,\rho_-+\delta\rho_-,\eta_B+\delta\eta_B]=\Psi[\rho_+^\dagger,\rho_-,\eta_B]
	\end{equation}
	relates $n$-point function to itself, $n$ being even, as
	\begin{equation}
		\langle\delta{\hat\zeta}_+({\bf x}_1){\hat\zeta}_-^\dagger({\bf x}_2)\cdots{\hat\zeta}_-^\dagger({\bf x}_n)\rangle + \langle{\hat\zeta}_+({\bf x}_1)\delta{\hat\zeta}_-^\dagger({\bf x}_2)\cdots{\hat\zeta}_-^\dagger({\bf x}_n)\rangle +
		\cdots+ \langle{\hat\zeta}_+({\bf x}_1){\hat\zeta}_-({\bf x}_2)\cdots\delta{\hat\zeta}_-^\dagger({\bf x}_n)\rangle=0
	\end{equation}
	where $\delta{\hat\zeta}_\pm({\bf x})$ is the transformation of primary Fermionic fields in CFT under conformal transformations. For more details, see appendix \ref{App.trandir}.

	\subsection{Ward Identity for U(1) Gauge Transformation}
	
	Finally, we explore the Ward identity from the invariance of wave function under $U(1)$ gauge transformation. Considering a $U(1)$ gauge field $A_{\mu}$ coupled with Fermionic field $\psi$ given by the action
	\begin{equation}
		\label{u1act}
		S={1\over2}\int d^{d+1}x \sqrt{-g} \bar{\psi}(ie^\mu_a\gamma^a{\cal D}_\mu-M)\psi
	\end{equation}
	where
	\begin{equation}
		{\cal D}_{\mu}=\nabla_\mu -ieA_\mu
	\end{equation}
	Under local $U(1)$ gauge transformation
	\begin{equation}
		\label{u1dif}
		\psi'(x)=e^{ie\chi(x)}\psi(x)~~,~~A'_\mu(x)=A_\mu(x)+\del_\mu\chi(x)
	\end{equation}
	One can find the action \eq{u1act} invariant. Working with the gauge choice
	\begin{equation}
		\label{a0gauge}
		A_0=0,
	\end{equation} 
	expansion of the wave functional in terms of the sources also includes contribution from gauge field as
	\begin{align}
		\log{\Psi}[\rho,A_i,\eta_B]&=\int d^dx_1d^dx_2 \sqrt{\hat g({\bf x}_1)}\sqrt{\hat g({\bf x}_2)} \rho^\dagger_+({\bf x}_1)\langle\zeta_+({\bf x}_1)\zeta_-^\dagger({\bf x}_2)\rangle \rho_-({\bf x}_2) \nonumber\\&+ \int d^dx_1d^dx_2d^dx_3 \sqrt{\hat g({\bf x}_1)}\sqrt{\hat g({\bf x}_2)}\sqrt{\hat g({\bf x}_3)}A_i({\bf x}_3)\rho_+^\dagger({\bf x}_1)\langle J^{i}({\bf x}_3)\zeta_+({\bf x}_1)\zeta_-^\dagger({\bf x}_2)\rangle \rho_-({\bf x}_2) + \cdots \label{PsigenA} 
	\end{align}
	Gauss's law implies that the wavefunctional above is invariant under the condition that the gauge parameter is independent on $\eta$ which preserves \eq{a0gauge}. Keeping therefore the parameter $\chi$ independent on $\eta$, 
	the infinitesimal $U(1)$ transformation is deduced from \eq{u1dif} as
	\begin{equation}
		\label{infu1}
		\delta\psi(x)=ie\chi({\bf x})\psi(x)~~,~~\delta A_i({\bf x})=\del_i\chi({\bf x})
	\end{equation}
	under which, one can easily find using \eq{rhopm_s} the transformation of $\rho_\pm$ as
	\begin{equation}
		\delta\rho_\pm({\bf x})=ie\chi({\bf x})\rho_\pm({\bf x})
	\end{equation}
	Invariance of the wave functional \eq{PsigenA} under the gauge transformation \eq{infu1},
	\begin{equation}
		\Psi[\rho_+^\dagger+\delta\rho_+^\dagger,\rho_-+\delta\rho_-,A_i+\delta A_i, \eta_B]=\Psi[\rho_+^\dagger,\rho_-,A_i,\eta_B]
	\end{equation}
	relates the correlation functions as
	\begin{equation}
		\del_i^{{\bf x}_3}\langle J^{i}({\bf x}_3){\hat\zeta}_+({\bf x}_1){\hat\zeta}_-^\dagger({\bf x}_2)\rangle+ie\delta({\bf x}_1-{\bf x}_3)\langle{\hat\zeta}_+({\bf x}_3){\hat\zeta}_-^\dagger({\bf x}_2)\rangle-ie\delta({\bf x}_2-{\bf x}_3)\langle{\hat\zeta}_+({\bf x}_1){\hat\zeta}_-^\dagger({\bf x}_3)\rangle=0
	\end{equation}
	This is indeed the Ward identity in CFT for U(1) Gauge transformation.
	
	As conclusion, we show that invariance of the wave functional under bulk and asymptotic isometries, as well as internal gauge symmetry, implies that the expansion coefficients in terms of appropriate source identification satisfy CFT Ward identities. It is expected to be true in more general circumstances, e.g. including colors etc, where one can even investigate for more Ward identities arising due to more gauge symmetries e.g. SU($N$). Considering the other representation \eq{psibypsial}, we can also get all the Ward identities discussed above with $\Delta_+$ replaced by $\Delta_-$, \eq{dim-} along with interchange of $+\leftrightarrow-$ in the subscript of the matrices and spinors. 

	\section{Discussions}
	\label{Sec.discuss}
	The study of de Sitter holography stands, as of now, at a premature stage compared to AdS/CFT correspondence. In fact, there exist two versions for the holographic correspondence in dS space -- one pertaining to static patch and the other is the late-time boundary. Considerable progress has been made in recent years in the understanding of dS space and its holographic aspects incorporating bosonic matter fields coupled to gravity. However, the Fermionic sector has received comparatively less attention in these studies\footnote{Some key references along this line include \cite{Fang:1979hq, Vasiliev:1987tk, Loran:2004fu, Pejhan:2019ech, Chen:2025foq}.}, to the best of our knowledge.
	In particular, unlike scalar fields, the de sitter invariant states, including Bunch Davies vacuum, for Fermionic theory are yet to be understood in great detail. Moreover one can also search for solutions of Wheeler de-Witt equation including Fermionic fields as the matter counter-part, possibly along the lines of \cite{Chakraborty:2023yed}. It is worth noting again that the derivation of the wavefunctional for fermionic states within the formalism of canonical quantization still lacks a concrete realisation, as stated in section \ref{sec.bdwf}.

	In this paper, we investigate some aspects of the dS/CFT correspondence pertaining to the late-time boundary version including Fermionic fields in spin-$1/2$ representation. In this version, the wavefunctional of a bulk state determined at late-time $\eta=\eta_B\ll 1$ is related to the partition function of the dual boundary theory. Based on a general understanding of holography, the late-time cut-off $\eta_B$ is related to the `UV cut-off' in the dual theory vanishing as $\eta_B\rightarrow0$.	
	
	Working in the expanding Poincare patch of dS space, we compute in the path integral approach the wave functional for the Bunch Davies vacuum in free Fermionic field theory in the basis constituted by the projections of Fermionic field \eq{splitpsi} onto the eigenspaces of $\gamma^0$. We show that a suitable pair of sources $\{{\hat\rho}_+,{\hat\rho}_-\}$ in the non-gravitating theory living on the boundary ${\cal I}^+$ can be identified, see \eq{rhopm_s} and \eq{rhohat}, such that the resulting wavefunction coefficient attains the form of the two point correlation function of a CFT.
	Correspondingly, a pair of boundary fields $\{{\hat\zeta}_+,{\hat\zeta}_-\}$ of conformal dimension $\{\Delta_+.\Delta_-\}$ is assigned in the dual CFT. Conformal invariance fixes the form of two point correlation function between these two dual fields as
	\begin{equation}
		\langle {\hat\zeta}_+({\bf x}){\hat\zeta}_-^\dagger({\bf y}) \rangle \sim {\gamma_{+-}^i (x_i-y_i)\over |{\bf x}-{\bf y}|^{2\Delta_+ +1}} ~~,~~ 		\langle {\hat\zeta}_-({\bf x}){\hat\zeta}_+^\dagger({\bf y}) \rangle \sim {\gamma_{-+}^i (x_i-y_i)\over |{\bf x}-{\bf y}|^{2\Delta_- +1}}
	\end{equation} 
	It turns out that when the wavefunctional is represented in terms of $\{\rho_+^\dagger,\rho_-\}$ (or, $\{\rho_-^\dagger,\rho_+\}$), the wavefunction coefficient takes the form of $\langle \zeta_+({\bf x})\zeta_-^\dagger({\bf y}) \rangle$ (or, $\langle \zeta_-({\bf x})\zeta_+^\dagger({\bf y}) \rangle$) mentioned above.  
	The conclusion is found to be robust under inclusion of Yukawa potential with scalar field belonging to both the complementary and principal series representation of the isometry group of dS space, see Appendix \ref{App.int}. We also analyse the OPE limit of the resulting three point correlation function which interestingly shows that taking the scalar field closer to either of the Fermionic insertions, ${\hat\zeta}_+$ or ${\hat\zeta}_-^\dagger$, it takes the form of two point correlator $\langle{\hat\zeta}_+({\bf x}){\hat\zeta}_-^\dagger({\bf x}')\rangle$, when sourced by $\{{\hat\rho}_+^\dagger,{\hat\rho}_-\}$. We show, given the identification \eq{rhopm_s}, that the invariance of the wavefunctional under spatial and time reparametrisations gives rise to the CFT Ward identities.
	
	Similarly, one can also
	 think of a hologram on ${\cal I}^-$, the boundary at far past and associate a pair of fields $\{{\hat{\tilde\zeta}}_+,{\hat{\tilde\zeta}}_-\}$ in the dual CFT sourced by $\{{\hat{\tilde\rho}}_+,{\hat{\tilde\rho}}_-\}$ defined as
	 	\begin{equation}
	 		\label{rhotildehat}
	 		{\tilde\rho}_-=\left[
	 		\begin{array}{c}
	 			0\\
	 			{\hat{\tilde\rho}}_{-}\\
	 		\end{array}
	 		\right]~~,~~
	 		{\tilde\rho}_+=\left[
	 		\begin{array}{c}
	 			{\hat{\tilde\rho}}_{+}\\
	 			0\\
	 		\end{array}
	 		\right]
	 	\end{equation}  
	 ${\tilde\rho}_+,{\tilde\rho}_-$ belonging to the positive and negative eigenspaces of $\gamma^0$ in the contracting Poincare patch ${\cal R}_-$, see Fig.\ref{poinpenrose}. Since ${\cal R}_-$ is time-reversed w.r.t. ${\cal R}_+$, the dimensions of $\{{\tilde\zeta}_+,{\tilde\zeta}_-\}$ are given by $\{\Delta_-,\Delta_+\}$ respectively. Schematically, the wavefunctional describing the patch ${\cal R}_-$ is given in the following representations
	 \begin{align}
	 	&\tilde{\Psi}[{\tilde\rho}_+^\dagger,{\tilde\rho}_-]=\exp\brt{\sum_{s_1s_2}\int d^dx_1d^dx_2 {\tilde\rho}^{s_1\dagger}_+({\bf x}_1)\langle{\tilde\zeta}_{+}({\bf x}_1){\tilde\zeta}_{-}^\dagger({\bf x}_2)\rangle {\tilde\rho}^{s_2}_-({\bf x}_2)} \label{wf+-lp}\\
	 	&\tilde{\Psi}[{\tilde\rho}_-^\dagger,{\tilde\rho}_+]=\exp\brt{\sum_{s_1s_2}\int d^dx_1d^dx_2 {\tilde\rho}^{s_1\dagger}_-({\bf x}_1)\langle{\tilde\zeta}_{-}({\bf x}_1){\tilde\zeta}_{+}^\dagger({\bf x}_2)\rangle {\tilde\rho}^{s_2}_+({\bf x}_2)} \label{wf-+lp}
	 \end{align}
	 where
	 \begin{equation}
	 	\langle {\tilde \zeta}_{+}({\bf x}){\tilde \zeta}_{-}^\dagger({\bf y}) \rangle \sim {\gamma_{+-}^i (x_i-y_i)\over |{\bf x}-{\bf y}|^{2\Delta_- +1}} ~~,~~ 		\langle {\tilde \zeta}_{-}({\bf x}){\tilde \zeta}_{+}^\dagger({\bf y}) \rangle \sim {\gamma_{-+}^i (x_i-y_i)\over |{\bf x}-{\bf y}|^{2\Delta_+ +1}}
	 \end{equation}
 	In Appendix \ref{App.corwfc}, we obtain in ${\cal R}_+$ the representation of the wavefunctional of Bunch Davies vacuum in the dual space using path integral technique. 
 	One can thereby also think of obtaining the wavefunctional in ${\cal R}_-$, \eq{wf+-lp} using the identification \eq{rhopm_s} in \eq{barpsi-+} at the late-time limit $\eta\rightarrow0$, similarly \eq{wf-+lp} from \eq{barpsi+-}, once ${\hat{\tilde\rho}}_\pm$ identified to ${\hat\rho}_\mp$. It is expected because the negative (or, positive) eigenspace of $\gamma^0$ in the expanding patch ${\cal R}_+$ turns to positive (or, negative) eigenspace of $\gamma^0$ in its time-reversed patch ${\cal R}_-$. In other words, perhaps pedagogically, the notions of Fermion and anti-Fermion reverse as we go from expanding to contracting Poincare patch. This argument is true, more generally, for any $\eta$-value in the bulk.

	It worths mentioning at this point that the following identification
	\begin{align}
		&\psi_-({\bf k},\eta_B)=-{2^{2iMR}e^{\pi MR}}{\Gamma[\nu_+]\over \Gamma[\nu_-]}(-\eta_B)^{{d\over2}-iMR}{\gamma^ik_i\over k}k^{-2iMR}\rho_+({\bf k}) \label{idal+}\\ 
		&\psi_+({\bf k},\eta_B)={2^{-2iMR}e^{\pi MR}}{\Gamma[\nu_-]\over \Gamma[\nu_+]}(-\eta_B)^{{d\over2}+iMR}{\gamma^ik_i\over k}k^{2iMR}\rho_-({\bf k}) \label{idal-}
	\end{align}  
	which, by putting in \eq{latewf}, give rise to the wavefunction coefficient \eq{wfcal} in the form of CFT correlator, instead of \eq{FFpos} that we obtained by using \eq{rhopm_s} in \eq{latewf}. The above identifications \eq{idal+}, \eqref{idal-} though certainly different from \eq{rhopm_s} are non-local in position space and therefore are not expected leading to Ward identities, as argued for scalar field in \cite{roy_dscft}. 

	Some progress has been done in the direction of reconstructing bulk in dS space \cite{Das:2012dt, Anninos:2011ui, Goldar:2024crc} including scalar fields and higher integer spins. It is also tempting to attempt the same with Fermionic fields, see \cite{Hertog:2019uhy}. As we noted earlier, the late-time boundary ${\cal I}^+$ is a Euclidean hypersurface and a proper understanding of dS holography is therefore expected to direct towards the understanding of emergence of time from the non-gravitating Euclidean dual theory. Our understanding of dS/CFT is very limited, no doubt, to resolve this mystery. But when we include Fermionic fields, we notice that the notion of $\gamma^0$, the time component of the $\gamma^a$, that exists in the bulk theory is not apparent in the boundary dual. For $d={\rm even}$, the boundary theory has the notion of chiral matrix, defined e.g. in $d=4$ as 
	\begin{equation}
		\gamma_{\rm ch}=\gamma^1\gamma^2\gamma^3\gamma^4
	\end{equation}
	having the following properties    
	\begin{equation}
		\{\gamma_{\rm ch},\gamma^i\}=0~~,~~ \gamma_{\rm ch}^\dagger=\gamma_{\rm ch} ~~,~~ \gamma_{\rm ch}^2={\mathbb I} ~~,~~ (\gamma^{i})^\dagger=\gamma_{\rm ch}\gamma^i\gamma_{\rm ch} ~~,~~i=1,2,3,4
	\end{equation}
	similar to $\gamma^0$ defined in the bulk. This suggests that the bulk notion of $\gamma^0$ is given by the boundary chiral matrix $\gamma_{\rm ch}$ \footnote{For any $d={\rm even}$, the definition takes the form,
		\begin{equation}
			\gamma_{ch}=(i\delta_{d,4n+2}+\delta_{d,4n})\gamma^1\gamma^2\gamma^3\gamma^4~,~n=0,1,2,\cdots.
		\end{equation}
	}, see \cite{Henning_ferads} for related discussion in Euclidean AdS. As a consequence, one can think of a field ${\zeta}$ on the boundary whose chiral projections will be dual to the  $\psi_+$ and $\psi_-$, i.e., the notions of Fermion and anti-Fermion, in the bulk. As argued in section \ref{sec.holo}, ${\hat\zeta}_\pm$ behave like primary Weyl spinors in the dual CFT and thereby the fields $\zeta_+, \zeta_-$ defined as
	\begin{equation}
	 		\label{zetaproj}
	 		{\zeta}_-=\left[
	 		\begin{array}{c}
	 			0\\
	 			{\hat{\zeta}}_{-}\\
	 		\end{array}
	 		\right]~~,~~
	 		{\zeta}_+=\left[
	 		\begin{array}{c}
	 			{\hat{\zeta}}_{+}\\
	 			0\\
	 		\end{array}
	 		\right]
	 	\end{equation}
can be assigned as the chiral projections of $\zeta$. However, since the conformal dimension of ${\hat\zeta}_{+}$ is different from ${\hat\zeta}_-$, the field $\zeta$ would not be a primary Dirac spinor in the dual theory. In the case for $d={\rm odd}$, ${\hat\zeta}_\pm$ behave as two independent primary Dirac spinors in the dual CFT as already discussed in section \ref{sec.holo}.
	
	Some comments about the correlation functions in the dual CFT need to be mentioned. In case of scalar fields, as discussed in \cite{roy_dscft}, the resulting late-time wave function coefficient after appropriate identification of sources, has a cut-off dependent local term in position space. The cut-off independent non-local term in position space in the coefficient takes the form of a CFT correlation function. In Fermionic theory, the wavefunction coefficient does not have any cut-off dependent term under the identification of sources \eq{rhopm_s}. Therefore, as time evolution in the bulk is unitary, an appropriate boundary CFT should encode all the information about the Bunch Davies vacuum state of Fermionic fields. 



	\section*{Acknowledgements}
	AR acknowledges insightful discussion with members of the TIFR String Theory Group, especially S.P. Trivedi, A. Gadde, S. Minwalla, O. Parrikar, S. Pal, S. Nath, P. Pandit, T. Chakraborty, I. Dey, S. Dhingra, K.S. Dhruva, H. Rajgadia and O. Nippanikar. AR also thanks A. Goldar, D. Majumder, R. Ghosh, D. Karan, K.K. Nanda and S. Sake for valuable discussion. AR acknowledges P. Dey, G. Senjor, and again S. Pal for sharing relevant references. AR acknowledges support from IISER Bhopal, CMI and IISER Pune to attend the workshops ``Asian Winter School", ``ST${}^4$" and ``Advances in Black Hole Theory" respectively while this article was under preparation. AR acknowledges support from the Department of Atomic Energy, The Government of India under the Project Identification Number RTI-4012, the Sudha Murthy Trust, and the Infosys Foundation. Finally, AR thanks the people of India for generously supporting research in String Theory.      
	
	\appendix
	
	\section{About Correlators and Wave Function Coefficient}
	\label{App.corwfc}
	In this appendix, we first canonically quantise the free Fermionic field in de Sitter space and compute the Wightman two point correlator w.r.t. Bunch Davies vacuum. Then we show that the correlator can be obtained by using the representations of the vacuum wavefunctional.
	
	The classical solution of Fermionic field in dS is given in \eq{psisol2dif} with $u,v$ given in \eq{usol} and \eq{vsol} respectively. Note that, near horizon, $\eta\rightarrow-\infty$,
	\begin{equation}
		u_s({\bf k},\eta)\sim e^{-ik\eta} ~~,~~v_s({\bf k},\eta)\sim e^{ik\eta}
	\end{equation}
	Identifying $u_s$ as the positive frequency solution, from the above equation, near the Poincare horizon, we quantise the Fermionic field as
	\begin{equation}
		\label{quanfer}
		{\psi}({\bf k},\eta)=\sum_s {\hat b}^s_{\bf k} N_u(k)u_s({\bf k},\eta)+({\hat c}^s_{\bf k})^{ \dagger} N_v(k)v_s({\bf k},\eta) 
	\end{equation}
	where ${\hat b}^s_{\bf k}$ and ${\hat c}^s_{\bf k}$ annihilate Bunch Davies vacuum $\ket{0}$
	\begin{equation}
		{\hat b}^s_{\bf k}\ket{0}=0~~,~~{\hat c}^s_{\bf k}\ket{0}=0
	\end{equation}
	and satisfy the following anti-commutation rules
	\begin{equation}
		\label{qrule}
		\{{\hat b}^s_{\bf k},({\hat b}^{s'}_{{\bf k}'})^{\dagger}\}=(2\pi)^d\delta({\bf k}-{\bf k}')\delta^{ss'}~~,~~\{{\hat c}^s_{\bf k},({\hat c}^{s'}_{{\bf k}'})^{\dagger}\}=(2\pi)^d\delta({\bf k}-{\bf k}')\delta^{ss'}
	\end{equation}
	and other possible anti-commutations between ${\hat b}^s_{\bf k},({\hat b}^{s}_{{\bf k}})^{\dagger},{\hat c}^s_{\bf k},({\hat c}^{s}_{{\bf k}})^{\dagger}$ are zero. The canonical momentum $\pi_\psi$ is given in \eq{mompsi}. Using \eq{quanfer}, we get
	\begin{equation}
		\pi_\psi({\bf k},\eta)={i\over2}R^d(-\eta)^{-d}\sum_s ({\hat b}^s_{\bf k})^{\dagger} N_u(k)^* u_s^\dagger({\bf k},\eta)+({\hat c}^s_{\bf k}) N_v(k)^* v_s^\dagger({\bf k},\eta) 
	\end{equation}
	Using the orthonormalisation conditions \eq{ortnorm}, it can be shown that the quantisation rule given in terms of the equal time anti-commutation as
	\begin{equation}
		\{{\psi}({\bf x},\eta),{\pi}_\psi({\bf x}',\eta)\}=i\delta({\bf x}-{\bf x}')
	\end{equation}
	fixes the normalisation constant $N_k$ as
	\begin{equation}
		N_u(k)=\sqrt{{\pi k}\over 4R^de^{\pi M R}}~~,~~ N_v(k)=\sqrt{{\pi k}\over 4R^de^{-\pi M R}}
	\end{equation}
	Using \eq{quanfer}, Wightman functions are calculated as
	\begin{align}
		&\bra{0}{\psi}({\bf k},\eta)\otimes{\psi}^\dagger({\bf k}',\eta)\ket{0}={{\pi  k}\over 4R^d e^{\pi M R}}\sum_s u_s({\bf k},\eta)u_s^\dagger({\bf k},\eta)\delta({\bf k}-{\bf k}')\label{uudag}\\
		&\bra{0}{\psi}^\dagger({\bf k},\eta)\otimes{\psi}({\bf k}',\eta)\ket{0}={{\pi  k}\over 4R^d e^{-\pi M R}}\sum_s v_s({\bf k},\eta)v_s^\dagger({\bf k},\eta)\delta({\bf k}-{\bf k}')\label{vvdag}
	\end{align}
	Using \eq{usol}, the \eq{uudag} reduces to
	\begin{align}
		&\bra{0}{\psi}({\bf k},\eta)\otimes{\psi}^\dagger({\bf k}',\eta)\ket{0}={{\pi  k}\over 4R^d e^{\pi M R}}\delta({\bf k}-{\bf k}')\times e^{\pi MR} \nonumber\\
		&\left[ \brf{X_{++}+{\gamma^ik_i\over k}X_{-+}} \sum_s a_-^s({\bf k})a_-^s({\bf k})^\dagger+ \brf{X_{--}+{\gamma^ik_i\over k}X_{+-}} \sum_s a_+^s({\bf k})a_+^s({\bf k})^\dagger\right] \label{psidagR}
	\end{align}
	where
	\begin{align}
		&X_{++}=(-\eta)^{1+d} e^{\pi MR}H^1_{\nu_-}(-k\eta)H^2_{\nu_+}(-k\eta)~~,~~X_{--}=(-\eta)^{1+d} e^{-\pi MR}H^1_{\nu_+}(-k\eta)H^2_{\nu_-}(-k\eta)\nonumber\\
		&X_{-+}= -(-\eta)^{1+d}H^2_{\nu_+}(-k\eta)H^1_{\nu_+}(-k\eta) ~~,~~ X_{+-}=(-\eta)^{1+d}H^2_{\nu_-}(-k\eta)H^1_{\nu_-}(-k\eta) \label{Xpm}
	\end{align}
	Noting that,
	\begin{equation}
		\sum_s a_-^s({\bf k})a_-^s({\bf k})^\dagger={1\over2}P_+ ~~,~~ \sum_s a_+^s({\bf k})a_+^s({\bf k})^\dagger={1\over2}P_-
	\end{equation}
	one can write using \eq{propPpm},
	\begin{align}
		&\bra{0}{\psi}({\bf k},\eta)\otimes{\psi}^\dagger({\bf k}',\eta)\ket{0}={{\pi  k}\over 8R^d }\delta({\bf k}-{\bf k}')\times \nonumber\\
		&\left[ \brf{P_+X_{++}P_+ + P_-{\gamma^ik_i\over k}X_{-+}P_+} + \brf{P_-X_{--}P_- + P_+{\gamma^ik_i\over k}X_{+-}P_-} \right] \label{psidagRP}
	\end{align}
	In Dirac basis, we can represent the above expression in block form as
	\begin{equation}
		\label{corrY}
		\bra{0}{\psi}({\bf k},\eta)\otimes{\psi}^\dagger({\bf k}',\eta)\ket{0}={{\pi  k}\over 8R^d }\delta({\bf k}-{\bf k}')\times \left[
		\begin{array}{cc}
			X_{++}{\mathbb I} & {\gamma_{+-}^ik_i\over k}X_{+-}  \\
			{\gamma_{-+}^ik_i\over k}X_{-+} & X_{--}{\mathbb I}  \\
		\end{array}
		\right]
	\end{equation}
	
	Similarly, using \eq{vsol}, the \eq{vvdag} reduces to
	\begin{align}
		\label{corrX}
		&\bra{0}{\psi}^\dagger({\bf k},\eta)\otimes{\psi}({\bf k}',\eta)\ket{0}={{\pi  k}\over 8R^d}\delta({\bf k}-{\bf k}')\times\left[
		\begin{array}{cc}
			X_{--}{\mathbb I} & -{\gamma_{+-}^ik_i\over k}X_{+-}  \\
			-{\gamma_{-+}^ik_i\over k}X_{-+} & X_{++}{\mathbb I}  \\
		\end{array}
		\right]
	\end{align}

	
	Now, we show that the quarter blocks in the above two expressions can be determined in path integral. 
To illustrate our argument, we first re-write all the representations of the wave functional \eq{psibypsich+-}, \eqref{psibypsich-+}, \eqref{psibypsich++}, \eqref{psibypsich--} as,
	\begin{equation}
		\label{psibypsire}
		\Psi_{+-}[{\psi_+^\dagger},\psi_-,\eta]=\exp\brt{-{R^d\over2}\sum_{s}\int {d^dk\over(2\pi)^d} (-{\eta})^{-d}{\hat\psi}_{+}^{s_1\dagger}({\bf k},{\eta}) {\mathbb M}^{-1}({\bf k},{\eta})_{+-}{\hat\psi}_{-}^{s_2}({\bf k},{\eta})}
	\end{equation}
	\begin{equation}
		\label{psibypsial1}
		\Psi_{++}[\psi_+^\dagger,\psi_+,\eta]=\exp\brt{-{R^d\over2}\sum_s \int{d^dk\over (2\pi)^d} (-\eta)^{-d} {\hat\psi}_+^{s_1\dagger}({\bf k},\eta)(1+{\mathbb M}^\dagger({\bf k},{\eta}) {\mathbb M}({\bf k},{\eta}))_{++}{\hat\psi}_+^{s_2}({\bf k},\eta)},
	\end{equation}
	\begin{equation}
		\label{psibypsial2}
		\Psi_{--}[\psi_-^\dagger,\psi_-,\eta]=\exp\brt{-{R^d\over2}\sum_s \int{d^dk\over (2\pi)^d} (-\eta)^{-d} {\hat\psi}_-^{s_1\dagger}({\bf k},\eta)(1+({\mathbb M}({\bf k},{\eta}){\mathbb M}^\dagger({\bf k},{\eta}))^{-1} )_{--}{\hat\psi}_-^{s_2}({\bf k},\eta)}
	\end{equation}
	and,
	\begin{equation}
		\label{psibypsial3}
		\Psi_{-+}[{\psi_-^\dagger},\psi_+,\eta]=\exp\brt{-{R^d\over2}\sum_{s}\int {d^dk\over(2\pi)^d} (-{\eta})^{-d}{\hat\psi}_{-}^{s_1\dagger}({\bf k},{\eta}) {{\mathbb M}({\bf k},{\eta})}_{-+}{\hat\psi}_{+}^{s_2}({\bf k},{\eta})}
	\end{equation}
	where ${\hat\psi}_\pm$ is defined as 
	\begin{equation}
		\label{psihats}
		{\psi}_-=\left[
		\begin{array}{c}
			0\\
			{\hat{\psi}}_{-}\\
		\end{array}
		\right]~~,~~
		{\psi}_+=\left[
		\begin{array}{c}
			{\hat{\psi}}_{+}\\
			0\\
		\end{array}
		\right]
	\end{equation}

	Second, in the dual space, we define the wavefunctional through path integral, compared to \eq{pifer}, as
	\begin{equation}
		\label{piferdag}
		{\bar\Psi}=\int d\psi d{\bar\psi} e^{-iS[\psi,{\bar\psi}]^{\dagger}}
	\end{equation}
	Noting that the complex conjugate of Dirac action \eq{frdiract} is given by
	\begin{equation}
		\label{sddag}
		S_D[\psi,{\bar\psi}]^{\dagger}=-{1\over2}\int d^{d+1}x \sqrt{-g}{\bar\psi}[i\overleftarrow{\nabla}_\mu e^\mu_a\gamma^a+M]\psi
	\end{equation}
	we see that specification of $\psi_-,\psi_+^\dagger$ at the boundary results the variation of \eq{sddag} to be
	\begin{equation}
		\delta (S_D[\psi,{\bar\psi}]^{\dagger})={i\over2} \int d^dx\sqrt{\gamma}(\delta\psi_-^\dagger\psi_-)
	\end{equation}
	By adding the following boundary term,
	\begin{equation}
		S_B^\dagger=-{i\over2}\int d^dx\sqrt{\gamma}(\psi_-^\dagger\chi_-)
	\end{equation}
	 and identifying $\chi_-=\psi_{-c}$,  we note that the variation of the total action $S_D^\dagger+S_B^\dagger$ vanishes at the saddle point. Using the total action in \eq{piferdag} as $S^\dagger=S_D^\dagger+S_B^\dagger$, we get at the saddle point,
	 \begin{equation}
	 	\label{barpsi+-}
	 	{\bar\Psi}_{+-}[\psi_+^\dagger,\psi_-]= \exp\brt{-{R^d\over2}\sum_s\int{d^dk\over(2\pi)^d}  (-\eta)^{-d} 
	{\hat\psi}_{+}^{s_1\dagger}({\bf k},{\eta}) {\mathbb M}^{\dagger}({\bf k},{\eta})_{+-}{\hat\psi}_{-}^{s_2}({\bf k},{\eta}) 	
 	 } 
	 \end{equation}
	Similarly, one can get other representations by specifying other three pairs of the projections, as mentioned in section \ref{sec.bdwf}, with addition of appropriate boundary terms $S_B^\dagger$ as
	\begin{align}
		\label{barpsi-+}
		{\bar\Psi}_{-+}[{\psi_-^\dagger},\psi_+,\eta]=\exp\brt{-{R^d\over2}\sum_{s}\int {d^dk\over(2\pi)^d} (-{\eta})^{-d}{\hat\psi}_{-}^{s_1\dagger}({\bf k},{\eta}) [{{\mathbb M}^{-1}({\bf k},{\eta})}]^\dagger_{-+}{\hat\psi}_{+}^{s_2}({\bf k},{\eta})}
	\end{align} 
	and,
	\begin{equation}
		{\bar\Psi}_{++}[\psi_+^\dagger,\psi_+,\eta]=\Psi_{++}[\psi_+^\dagger,\psi_+,\eta] ~~,~~ {\bar\Psi}_{--}[\psi_-^\dagger,\psi_-,\eta]=\Psi_{--}[\psi_-^\dagger,\psi_-,\eta]
	\end{equation}
	The expectation value of the operators in path integral is computed as
	\begin{equation}
		\langle{\hat\psi}_b({\bf k}',\eta){\hat\psi}^{\dagger}_a({\bf k},\eta)\rangle \equiv \int {\cal D}{\hat\psi}_a^\dagger{\cal D}{\hat\psi}_b {\bar\Psi}_{ab}[\psi_a^\dagger,\psi_b,\eta] {\hat\psi}_b({\bf k}',\eta){\hat\psi}_a^{\dagger}({\bf k},\eta) \Psi_{ab}[\psi_a^\dagger,\psi_b,\eta] 
	\end{equation}
	where the indices $a,b$ assume the signs $\{+,-\}$. 
	    
	Finally,
	as $\psi_\pm,\psi_\pm^\dagger$ are Grassmann variables, the above integral results
	\begin{align}
		&\langle{\hat\psi}_+({\bf k}',\eta){\hat\psi}^{\dagger}_+({\bf k},\eta)\rangle = {-R^{-d}\over (-\eta)^{-d}} ((1+{\mathbb M}^\dagger{\mathbb M})_{++})^{-1}\delta({\bf k}-{\bf k}')={-R^{-d}\over (-\eta)^{-d}}{H^2_{\nu_-}(-k\eta)H^1_{\nu_+}(-k\eta)\over Q(k,\eta)}{\mathbb I}\label{+dag+}\\
		&\langle{\hat\psi}_-({\bf k}',\eta){\hat\psi}^{\dagger}_-({\bf k},\eta)\rangle = {-R^{-d}\over (-\eta)^{-d}} ((1+({\mathbb M}{\mathbb M}^\dagger)^{-1})_{--})^{-1}\delta({\bf k}-{\bf k}')={-R^{-d}\over (-\eta)^{-d}}{e^{2\pi MR}H^2_{\nu_+}(-k\eta)H^1_{\nu_-}(-k\eta)\over Q(k,\eta)}{\mathbb I} \label{-dag-}\\
		&\langle{\hat\psi}_-({\bf k}',\eta){\hat\psi}^{\dagger}_+({\bf k},\eta)\rangle = {-2R^{-d}\over (-\eta)^{-d}} (({\mathbb M}^\dagger+{\mathbb M}^{-1})_{+-})^{-1}\delta({\bf k}-{\bf k}')=-{2R^{-d}\over (-\eta)^{-d}}{e^{\pi MR}H^2_{\nu_+}(-k\eta)H^1_{\nu_+}(-k\eta)\over Q(k,\eta)}{\gamma_{-+}^i k_i\over k} \label{+dag-}\\
		&\langle{\hat\psi}_+({\bf k}',\eta){\hat\psi}^{\dagger}_-({\bf k},\eta)\rangle = {-2R^{-d}\over (-\eta)^{-d}} ((({\mathbb M}^\dagger)^{-1}+{\mathbb M})_{-+})^{-1}\delta({\bf k}-{\bf k}')={2R^{-d}\over (-\eta)^{-d}}{e^{\pi MR}H^2_{\nu_-}(-k\eta)H^1_{\nu_-}(-k\eta)\over Q(k,\eta)}{\gamma_{+-}^ik_i\over k} \label{-dag+}
	\end{align}
	where $Q(k,\eta)$ is given by
	\begin{equation}
		Q(k,\eta)\equiv H^2_{\nu_-}(-k\eta)H^1_{\nu_+}(-k\eta)+e^{2\pi MR}H^2_{\nu_+}(-k\eta)H^1_{\nu_-}(-k\eta)=-{4e^{\pi MR}\over \pi k\eta}
	\end{equation}
	where we use \eq{nupm} and the Wronskian formula of Hankel functions to derive the above equality. Note that we have also used \eq{+-inv} in \eq{+dag-}, \eq{-dag+}. Comparing \eq{+dag+}--\eqref{-dag+} with \eq{Xpm}, we get
	\begin{equation}
		\label{expcoreq}
		\langle{\hat\psi}_b({\bf k}',\eta){\hat\psi}^\dagger_a({\bf k},\eta)\rangle =-{{\pi  k}\over 4R^d}\delta({\bf k}-{\bf k}')\times\begin{cases}
			-2{\gamma_{ba}^i k_i \over k}X_{ba} & \text{if } a \neq b \\
			X_{{\bar a}{\bar b}} & \text{if } a = b
		\end{cases}
	\end{equation}
	where ${\bar a}, {\bar b}$ assume the opposite of sign taken by $a,b$ respectively. Comparing the above result with \eq{corrY} and \eq{corrX}, we get
	\begin{equation}
		\bra{0}{\psi}({\bf k},\eta)\otimes{\psi}^\dagger({\bf k}',\eta)\ket{0}=-{1\over2}\times \left[
		\begin{array}{cc}
			\langle{\hat\psi}_-({\bf k}',\eta){\hat\psi}^\dagger_-({\bf k},\eta)\rangle & -{1\over2}\langle{\hat\psi}_+({\bf k}',\eta){\hat\psi}^\dagger_-({\bf k},\eta)\rangle  \\
			-{1\over2}\langle{\hat\psi}_-({\bf k}',\eta){\hat\psi}^\dagger_+({\bf k},\eta)\rangle & \langle{\hat\psi}_+({\bf k}',\eta){\hat\psi}^\dagger_+({\bf k},\eta)\rangle  \\
		\end{array}
		\right]
	\end{equation}
	\begin{equation}
		\bra{0}{\psi}^\dagger({\bf k}',\eta)\otimes{\psi}({\bf k},\eta)\ket{0}=-{1\over2}\times \left[
		\begin{array}{cc}
			\langle{\hat\psi}_+({\bf k}',\eta){\hat\psi}^\dagger_+({\bf k},\eta)\rangle & {1\over2}\langle{\hat\psi}_+({\bf k}',\eta){\hat\psi}^\dagger_-({\bf k},\eta)\rangle  \\
			{1\over2}\langle{\hat\psi}_-({\bf k}',\eta){\hat\psi}^\dagger_+({\bf k},\eta)\rangle & \langle{\hat\psi}_-({\bf k}',\eta){\hat\psi}^\dagger_-({\bf k},\eta)\rangle  \\
		\end{array}
		\right]
	\end{equation}
	Therefore, we see that the two point correlator of Fermionic operators obtained from canonical approach can also be reproduced by computing the two point expectation values of its projections 
	as mentioned in section \ref{sec.bdwf}.

	\section{Representations in Different Space-time Dimensions}
	\label{App.subtled}
	In this appendix, we present some of the representations of $\gamma$-matrices in 3, 4 and 5 dimensional manifolds. In particular, we also note the eigenvectors of $\gamma^0$ in 4D, as promised in section \ref{sec.holo}. At the end, we will comment on the `quarter blocks' of $\gamma^ik_i\over k$. Standard references include \cite{Thaller:1992fcx, Angelone:2022tet, Lonigro:2022fac}. 
	


	\paragraph{Euclidean 3D}
	On 3D Euclidean manifold, the irreducible representations of $\gamma$-matrices satisfying the properties
	\begin{equation}
		\label{eugamal}
		\{\gamma^i,\gamma^j\}=-2\delta_{ij}{\mathbb I}~~,~~(\gamma^i)^\dagger=-\gamma^i
	\end{equation}
	are given by
	\begin{equation}
		\label{3dgam}
		\gamma^1=i\left(
		\begin{array}{cc}
			{0} & {1} \\
			{1} & {0} \\
		\end{array}
		\right)~~,~~
		\gamma^2=i\left(
		\begin{array}{cc}
			{0} & {-i} \\
			{i} & {0} \\
		\end{array}
		\right)~~,~~
		\gamma^3=i\left(
		\begin{array}{cc}
			{1} & {0} \\
			{0} & {-1} \\
		\end{array}
		\right)
	\end{equation}
	Note that the $\gamma^3$ in Euclidean 3D assumes the Dirac form of $\gamma^0$ in Lorentzian 3D manifold, multiplied by a factor of $i$ due to appropriate Hermiticity conditions.  	
	\paragraph{Lorentzian 4D}	In 4D Lorentzian manifold, representation of $\gamma$-matrices, satisfying the properties \eq{gacomdag}, in Dirac basis are given by
		\begin{align}
			&\gamma^0=\left[
			\begin{array}{cccc}
				1 & 0 & 0 & 0 \\
				0 & 1 & 0 & 0 \\
				0 & 0 & -1 & 0 \\
				0 & 0 & 0 & -1 \\
			\end{array}
			\right]~~,~~
			\gamma^1=\left[
			\begin{array}{cccc}
				0 & 0 & 0 & 1 \\
				0 & 0 & 1 & 0 \\
				0 & -1 & 0 & 0 \\
				-1 & 0 & 0 & 0 \\
			\end{array}
			\right] \nonumber\\
			&\gamma^2=\left[
			\begin{array}{cccc}
				0 & 0 & 0 & -i \\
				0 & 0 & i & 0 \\
				0 & i & 0 & 0 \\
				-i & 0 & 0 & 0 \\
			\end{array}
			\right]~~,~~
			\gamma^3=\left[
			\begin{array}{cccc}
				0 & 0 & 1 & 0 \\
				0 & 0 & 0 & -1 \\
				-1 & 0 & 0 & 0 \\
				0 & 1 & 0 & 0 \\
			\end{array}
			\right] \label{dirbasis}
		\end{align}
		The set of spinors $a^s_\pm$ satisfying \eq{g0eig}, \eq{ortnorm} and \eq{a+a-rel} are given as
		\begin{equation}
			a^1_+=(0,0,{1\over\sqrt{2}},0)~,~a^2_+=(0,0,0,{1\over\sqrt{2}})~,~a^1_-=({k_3\over \sqrt{2}k},{k_+\over \sqrt{2}k},0,0)~,~a^2_-=({k_-\over \sqrt{2}k},-{k_3\over \sqrt{2}k},0,0)
		\end{equation}
		where $k_\pm\equiv k_1\pm ik_2$. Note that the $\gamma^i$ in \eq{3dgam} takes the form of the top-right quarter blocks of respective $\gamma^i$ in \eq{dirbasis}, multipled by factor of $i$ due to Hermiticity conditions. 
		
		In Weyl basis, on the other hand, $\gamma^0$ takes the form
		\begin{equation}
			\label{weyl4}
			\gamma^0=\left[
			\begin{array}{cccc}
				0 & 0 & 1 & 0 \\
				0 & 0 & 0 & 1 \\
				1 & 0 & 0 & 0 \\
				0 & 1 & 0 & 0 \\
			\end{array}
			\right]
		\end{equation}
		while $\gamma^i$ remains in the same form as in \eq{dirbasis}. Corresponding $a_\pm^s$ are given as
		\begin{equation}
			a^1_+=(0,-{1\over 2},0,{1\over 2})~,~a^2_+=(-{1\over 2},0,{1\over 2},0)~,~a^1_-=({k_-\over 2k},-{k_3\over 2k},{k_-\over 2k},-{k_3\over 2k})~,~a^2_-=({k_3\over 2k},{k_+\over 2k},{k_3\over 2k},{k_+\over 2k})
		\end{equation}
%

		\paragraph{Lorentzian 5D}
		
		In 5D Lorentzian manifold, representation of $\gamma$-matrices, satisfying the properties \eq{gacomdag}, in Dirac basis are given by
		\begin{align}
			&\gamma^0=\left[
			\begin{array}{cccc}
				1 & 0 & 0 & 0 \\
				0 & 1 & 0 & 0 \\
				0 & 0 & -1 & 0 \\
				0 & 0 & 0 & -1 \\
			\end{array}
			\right]~~,~~
			\gamma^1=\left[
			\begin{array}{cccc}
				0 & 0 & 0 & 1 \\
				0 & 0 & 1 & 0 \\
				0 & -1 & 0 & 0 \\
				-1 & 0 & 0 & 0 \\
			\end{array}
			\right] \nonumber\\
			&\gamma^2=\left[
			\begin{array}{cccc}
				0 & 0 & 0 & -i \\
				0 & 0 & i & 0 \\
				0 & i & 0 & 0 \\
				-i & 0 & 0 & 0 \\
			\end{array}
			\right]~~,~~
			\gamma^3=\left[
			\begin{array}{cccc}
				0 & 0 & 1 & 0 \\
				0 & 0 & 0 & -1 \\
				-1 & 0 & 0 & 0 \\
				0 & 1 & 0 & 0 \\
			\end{array}
			\right] ~~,~~
				\gamma^4=i\left[
			\begin{array}{cccc}
					0 & 0 & 1 & 0 \\
				0 & 0 & 0 & 1 \\
				1 & 0 & 0 & 0 \\
				0 & 1 & 0 & 0 \\
			\end{array}
			\right]\label{dirbasis5}
		\end{align}
		Note that the $\gamma^i$ in the Lorentzian 5D are all of the $\gamma$-matrices in Euclidean 4D manifold satisfying \eq{eugamal}. Also, note that the $\gamma^4$ in Euclidean 4D assumes the Weyl form of $\gamma^0$ in Lorentzian 4D manifold, \eq{weyl4}, multiplied by a factor of $i$ due to appropriate Hermiticity conditions.

	 As we discussed in section \ref{sec.holo}, the correlation functions of the boundary fields take a matrix form due to the factor ${\gamma_{\pm\mp}^ik_i\over k}$ where $\gamma_{\pm\mp}^i$ are the off-diagonal blocks of $\gamma^i$ defined in the bulk. Let us now consider dS${}_4$/CFT${}_3$ holography where ${\gamma^ik_i\over k}$ in the bulk takes the form in Dirac basis as 
	\begin{equation}
		\label{gamk4}
		{\gamma^i k_i\over k}={1\over k}\left(
		\begin{array}{cccc}
			0 & 0 & {k_3} & {k_1}-i {k_2} \\
			0 & 0 & {k_1}+i {k_2} & -{k_3} \\
			-{k_3} & -{k_1}+i {k_2} & 0 & 0 \\
			-{k_1}-i {k_2} & {k_3} & 0 & 0 \\
		\end{array}
		\right)
	\end{equation}
		On the other hand, using the IrReps of $\gamma$ matrices, \eq{3dgam}, defined on the manifold of ${\cal I}^+$, we get
		\begin{equation}
			\label{gamk3dirp}
			{\gamma^i k_i\over k}={i\over k}\left(
			\begin{array}{cc}
				{k_3} & {k_1}-i {k_2} \\
				{k_1}+i {k_2} & -{k_3} \\
			\end{array}
			\right)
		\end{equation}
%
	which is indeed any of the off-diagonal blocks of \eq{gamk4}, upto a factor of $\pm i$.

In dS${}_5$/CFT${}_4$ holography, ${\gamma^ik_i\over k}$ in the bulk takes the form in Dirac basis as 
\begin{equation}
	{\gamma^i k_i\over k}={1\over k}\left(
	\begin{array}{cccc}
		0 & 0 & {k_3}+i {k_4} & {k_1}-i {k_2} \\
		0 & 0 & {k_1}+i {k_2} & -{k_3}+i {k_4} \\
		-{k_3}+i {k_4} & -{k_1}+i {k_2} & 0 & 0 \\
		-{k_1}-i {k_2} & {k_3}+i {k_4} & 0 & 0 \\
	\end{array}
	\right)
\end{equation}
It is to note that in both the cases, more generally in any dimensions, the Dirac representation of ${\gamma^ik_i\over k}$ in the Lorentzian bulk carries only the off-diagonal quarter blocks ${\gamma_{\pm\mp}^ik_i\over k}$. This implies,
\begin{equation}
	\label{+-inv}
	\brf{\gamma_{+-}^ik_i\over k}^{-1}=-\brf{\gamma_{-+}^ik_i\over k}
\end{equation} 

		\section{Transformation of Fermionic Field}
		\label{App.trandir}
		In this appendix, we provide more details about change in Fermionic fields under both infinitesimal local Lorentz transformation (ILLT) and diffeomorphisms. 
		\subsection{Transformation under ILLT}
		Following the introduction of n-bein formalism in section \ref{sec.clsol}, n-bein is uniquely determined by spacetime metric up to a local Lorentz transformation (LLT) represented by matrix $\Lambda$. More precisely, performing pointwise independent Lorentz transformation on each of the local orthonormal frames, the change in the n-bein is given as    
		\begin{equation}
			\label{etransLLT}
			e'^a_\mu(x)=\Lambda^a_b(x)e^b_\mu(x)
		\end{equation}
		Since $\Lambda$ satisfies
		\begin{equation}
			\label{lamlameta}
			\Lambda^a_c\Lambda^b_d\eta_{ab}=\eta_{cd}
		\end{equation}
		one can re-write \eq{etransLLT} as
		\begin{equation}
			\label{etrinvllt}
			e'^\mu_a(x)=(\Lambda^{-1}(x))^b_a e^\mu_b(x)
		\end{equation}
		
		Let us consider $\psi$ to be a Dirac field which transforms under local Lorentz transformation, represented by matrix $\Lambda$, as
		\begin{equation}
			\label{dirtrllt}
			\psi'(x)=S(\Lambda(x))\psi(x)
		\end{equation}
		where $S(\Lambda)$ denotes a representation of the Lorentz group. In subsequent discussion, we will see that $S(\Lambda)$ turns out a representation of the double covering of the Lorentz group, see \cite{Iliesiu:2015qra} for more discussions in 3 dimensions.
		The covariant derivative of $\psi$ should therefore transform under LLT as
		\begin{equation}
			\label{nabdirtrllt}
			\nabla_\mu\psi'(x)=S(\Lambda(x))\nabla_\nu\psi(x)
		\end{equation}
		Using \eq{etrinvllt}, \eqref{dirtrllt}, \eqref{nabdirtrllt}, LHS of Dirac equation \eq{freomdirac} transforms under the LLT as
		\begin{equation}
			[ie^\mu_a(x)\gamma^a\nabla_\mu-M]\psi\rightarrow[i\Lambda^b_a(x)e'^\nu_b(x)\gamma^aS^{-1}(\Lambda(x))\nabla_\nu-MS^{-1}(\Lambda(x))]\psi'(x)
		\end{equation}
		By virtue of \eq{freomdirac}, we can write
		\begin{equation}
			[i\Lambda^b_a(x)e'^\nu_b(x) S(\Lambda(x))\gamma^aS^{-1}(\Lambda(x))\nabla_\nu-M]\psi'(x)=0
		\end{equation}
		Invariance of Dirac equation under the LLT essentially implies
		\begin{equation}
			\label{dirinvcond}
			\Lambda^b_a(x) S(\Lambda(x))\gamma^aS^{-1}(\Lambda(x))=\gamma^b
		\end{equation}
		Characterizing an infinitesimal local Lorentz transformation (ILLT) as
		\begin{equation}
			\label{ILLT}
			\Lambda^a_b(x)=\delta^a_b+\epsilon^a_b(x)
		\end{equation}
		$S(\Lambda)$ can be written under \eq{ILLT} as
		\begin{equation}
			\label{infSlam}
			S(\Lambda(x)=1+\epsilon(x))= 1+i\epsilon^{ab}(x){\cal M}_{ab}
		\end{equation}
		where ${\cal M}_{ab}$ stands for a representation of $so(d,1)$, the Lie algebra of the Lorentz group $SO(d,1)$. Note that since $\epsilon_{ab}$ is anti-symmetric due to \eq{lamlameta},
		${\cal M}_{ab}$ is anti-symmetric in indices. 
		
		The infinitesimal version of \eq{dirinvcond} can be derived using \eq{ILLT}, \eqref{infSlam} as
		\begin{equation}
			\epsilon^b_a(x)\gamma^a+i\epsilon_{cd}[{\cal M}^{cd},\gamma^b]=0
		\end{equation}
		One can check that the above condition is satisfied with
		\begin{equation}
			{\cal M}^{ab}={i\over 8}[\gamma^a,\gamma^b]
		\end{equation}
		Hence under ILLT, we get from \eq{dirtrllt}
		\begin{equation}
			\psi'(x)=(1-{1\over 8}\epsilon_{ab}(x)[\gamma^a,\gamma^b])\psi(x)
		\end{equation}
		as mentioned in \eq{deldirillt}.

		\subsection{Transformation under Diffeomorphisms}
		In section \ref{Sec.WI}, we discuss the transformation of Fermionic field, \eq{gendelwf} under diffeomorphism given in \eq{genco}. Here we provide additional details, typically tied to deriving \eq{delwfsp}, \eq{delwftime}, and transformations under the action of conformal generators.
		
		We first re-write \eq{gendelwf} as follows,
		\begin{equation}
			\delta\psi(x)\equiv\psi'(x)-\psi(x)=-\xi^\mu\nabla_\mu\psi(x)+{1\over8}\nabla_\mu\xi_\nu e^\mu_a e^\nu_b[\gamma^a,\gamma^b]\psi(x)
		\end{equation}  
		Using \eq{covDfermion} and \eq{gtoeta}, above expression reduces to
		\begin{equation}
			\label{delpsiexp2}
			\delta\psi(x)=-\xi^\mu\del_\mu\psi(x)+{1\over 8}\xi^\mu\omega_{\mu ab}[\gamma^a,\gamma^b]\psi(x)+{1\over8}\eta_{bc}e^c_\rho e^\mu_a(\del_\mu\xi^\rho+\Gamma^\rho_{\mu\alpha}\xi^\alpha)[\gamma^a,\gamma^b]\psi(x)
		\end{equation}
		Using the expressions given in \eq{nbeinds}, \eq{cristds} and \eq{spinconds}, we get
		\begin{equation}
			(\xi^\mu\omega_{\mu ab}+\eta_{bc}e^c_\rho e^\mu_a \Gamma^\rho_{\mu\alpha}\xi^\alpha)[\gamma^a,\gamma^b]=0
		\end{equation}
		Hence, \eq{delpsiexp2} reduces to
		\begin{equation}
			\label{delpsi}
			\delta\psi(x)=-\xi^\mu\del_\mu\psi(x)+{1\over8}\eta_{b\rho} \del_a\xi^\rho[\gamma^a,\gamma^b]\psi(x)
		\end{equation}
		Note that, for spatial reparametrisation given by \eq{spdif}, the above expression can be readily reduced to \eq{delwfsp}. Similarly for time reparametrisation given by \eq{trep}, one can reproduce \eq{delwftime} from the above expression.

		\paragraph{Conformal transformations}
		Conformal transformations are generated due to the diffeomorphisms mentioned in \eq{trdif}--\eqref{sctdif}. One can calculate the transformation of Fermionic field under these diffeomorphisms using \eq{delpsi} as follows
		\begin{align}
			&{\rm Translation}:~~~~\delta\psi(x)=-s^i\del_i\psi(x)\\
			&{\rm Rotation} : ~~~~ \delta\psi(x)=-s^i_{{~}j} x^j\del_i\psi(x)-{1\over8}s_{ij}[\gamma^i,\gamma^j]\psi(x)\\
			&{\rm Dilatation} :~~~~ \delta\psi(x)=-s(x^i\del_i\psi+\eta\del_\eta\psi)\\
			&{\rm SCT}:~~~~ \delta\psi(x)= -[(2x^ib_jx^j-x^jx_jb^i)\del_i\psi+2b^jx_j\eta\del_\eta\psi]+{1\over 2}b_ix_j[\gamma^i,\gamma^j]\psi
		\end{align}
		Correspondingly, the source $\rho_\pm({\bf x})$ defined by \eq{rhopm_s} transforms as
		\begin{align}
			&{\rm Translation}:~~\delta\rho_\pm({\bf x})=-s^i\del_i\rho_\pm({\bf x})\label{trrho}\\
			&{\rm Rotation} : ~~ \delta\rho_\pm({\bf x})=-s^i_{{~}j} x^j\del_i\rho_\pm({\bf x})-{1\over8}s_{ij}[\gamma^i,\gamma^j]\rho_\pm({\bf x})\\
			&{\rm Dilatation} :~~ \delta\rho_\pm({\bf x})=-s(x^i\del_i\rho_\pm({\bf x})+\Delta_\pm\rho_\pm({\bf x}))\\
			&{\rm SCT}:~~ \delta\rho_\pm({\bf x})= -[(2x^ib_jx^j-x^jx_jb^i)\del_i\rho_\pm({\bf x})+2b^jx_j\Delta_\pm\rho_\pm({\bf x})]+{1\over 2}b_ix_j[\gamma^i,\gamma^j]\rho_\pm({\bf x})\label{sctrho}
		\end{align}
		As we have discussed in section \ref{sec.holo}, $\rho_\pm({\bf x})$ is identified as the source of the boundary field theory. The transformations of the source $\rho_\pm$ under the conformal diffeomorphisms generate changes in the partition function of the boundary field theory. Noting that the late-time wavefunctional in the bulk is equal to the partition function of the boundary theory, the change in the free theory wave functional is given by
		\begin{align}
			\delta\log\Psi_{FT}= &\sum_s\int d^dx_1d^dx_2 \delta \rho^{s_1\dagger}_+({\bf x}_1)\langle\zeta_{+}({\bf x}_1)\zeta_{-}^\dagger({\bf x}_2)\rangle \rho^{s_2}_-({\bf x}_2)\nonumber\\&+\sum_s\int d^dx_1d^dx_2 \rho^{s_1\dagger}_+({\bf x}_1)\langle\zeta_{+}({\bf x}_1)\zeta_{-}^\dagger({\bf x}_2)\rangle \delta\rho^{s_2}_-({\bf x}_2)
		\end{align}
		The change due to transformation of the source $\rho_\pm$ can also be understood as the change in the wavefunction coefficient $\langle\zeta_{+}({\bf x}_1)\zeta_{-}^\dagger({\bf x}_2)\rangle$. From boundary perspective, the change in the coefficient originates due to the transformation of the boundary fields ${\hat\zeta}_\pm({\bf x})$ as,
		\begin{align}
			\delta\log\Psi_{FT}= &\sum_s\int d^dx_1d^dx_2  {\hat\rho}^{s_1\dagger}_+({\bf x}_1)\langle \delta{\hat\zeta}_{+}({\bf x}_1){\hat\zeta}_{-}^\dagger({\bf x}_2)\rangle {\hat\rho}^{s_2}_-({\bf x}_2)\nonumber\\&+\sum_s\int d^dx_1d^dx_2 {\hat\rho}^{s_1\dagger}_+({\bf x}_1)\langle{\hat\zeta}_{+}({\bf x}_1)\delta{\hat\zeta}_{-}^\dagger({\bf x}_2)\rangle {\hat\rho}^{s_2}_-({\bf x}_2)
		\end{align}
		where, the transformation of ${\hat\zeta}_\pm({\bf x})$ is obtained from the transformation formulae of the source $\rho_\pm$, \eq{trrho}--\eqref{sctrho} as,
		\begin{align}
			&{\rm Translation}:~~\delta{\hat\zeta}_\pm({\bf x})=s^i\del_i{\hat\zeta}_\pm({\bf x})\label{trbf}\\
			&{\rm Rotation} : ~~ \delta{\hat\zeta}_\pm({\bf x})=s^i_{{~}j} x^j\del_i{\hat\zeta}_\pm({\bf x})+{1\over8}s_{ij}[\gamma^i,\gamma^j]_{\pm\pm}{\hat\zeta}_\pm({\bf x})\\
			&{\rm Dilatation} :~~ \delta{\hat\zeta}_\pm({\bf x})=s(x^i\del_i{\hat\zeta}_\pm({\bf x})+(d-\Delta_\mp){\hat\zeta}_\pm({\bf x}))\label{dilzeta}\\
			&{\rm SCT}:~~ \delta{\hat\zeta}_\pm({\bf x})= [(2x^ib_jx^j-x^jx_jb^i)\del_i{\hat\zeta}_\pm({\bf x})+2b^jx_j(d-\Delta_\mp){\hat\zeta}_\pm({\bf x})]-{1\over 2}b_ix_j[\gamma^i,\gamma^j]_{\pm\pm}{\hat\zeta}_\pm({\bf x})\label{sctbf}
		\end{align}
		which take part in the conformal Ward identities, as discussed in Section \ref{sec.WIconf}.
		From the above expressions, one can show that ${\hat\zeta}_\pm({\bf x})$ indeed behave like a primary field of a fermionic CFT with conformal dimension $\Delta_\pm$ as mentioned in section \ref{sec.holo}. Similar expressions for primary spinor fields are illustrated in \cite{Iliesiu:2015qra}. More explicitly, one can associate operators that generate the above conformal transformations. The representations of these operators when acting on the boundary Fermionic field ${\hat\zeta}_\pm({\bf x})$ are as follows :  
		\begin{align}
			&{\rm Translation}:~~P_i=i\del_i\\
			&{\rm Rotation} : ~~ M_{ij}=i[ (x^j\del_i-x^i\del_j)+{1\over4}[\gamma^i,\gamma^j]_{\pm\pm}]\\
			&{\rm Dilatation} :~~ D=i(x^i\del_i+(d-\Delta_\mp))\\
			&{\rm SCT}:~~ K_i= i[2x_ix^j\del_j-x^jx_j\del_i+2x_i(d-\Delta_{\mp})-{1\over2}x_j[\gamma^i,\gamma^j]_{\pm\pm}]
		\end{align}
		One can easily check that the above generators satisfy the algebra of a conformal group \cite{DiFrancesco:1997nk, Qualls:2015qjb}.
		
		\section{Interaction with Scalar Field}
		\label{App.int}
		In section \ref{sec.holo}, we proposed an identification of source in terms of the projections of Fermionic fields, and showed in free theory, the wavefunction coefficient of BD vacuum, under this identification, takes the form of a CFT two point function. In this appendix, we study the robustness of the identification by including Yukawa interaction term in the action at first order of the coupling constant. We set $R=1$ for the following discussions.     
		\subsection{Review of Identifications of Sources for Scalar Field}
		\label{app.scalar}
		Representation of scalar fields of mass $M_0$ in de Sitter space lies in complementary series for $M_0<d/2$, which we identify as `overdamped scalars', and principal series for $M_0>d/2$, identified as `underdamped scalars'. Holographic dictionaries for these two series of representation are different in terms of identification of sources as given in \cite{roy_dscft}, and we present a brief review of the related discussions. 
		
		Considering a free scalar theory in dS space,
		\begin{equation}
			S=-{1\over2}\int d^{d+1}x \sqrt{-g}(g^{\mu\nu}\del_\mu\phi\del_\nu\phi+M_0^2\phi^2) 
		\end{equation}
		the classical solution in Poincare coordinates is given by
		\begin{equation}
			\phi({\bf x},\eta)=\int {d^dk\over (2\pi)^d}[C_1({\bf k})(-\eta)^{d/2}H^1_\nu(-k\eta)+C_2({\bf k})(-\eta)^{d/2}H^2_\nu(-k\eta)]e^{i{\bf k}.{\bf x}}
		\end{equation}
		where $\nu$ is given by
		\begin{equation}
			\nu=\sqrt{{d^2\over4}-M_0^2}
		\end{equation}
		For overdamped case, $\nu$ is real; for underdamped case, $\nu$ is imaginary and for convenience is written as $\nu=i\mu$, where $\mu$ is real. We define
		\begin{equation}
			F_\nu(k,\eta)\equiv C_\nu(-\eta)^{d/2}H^1_\nu(-k\eta)~~,~~F_\mu(k,\eta)\equiv C_\mu (-\eta)^{d/2}H^1_{i\mu}(-k\eta)
		\end{equation}
		where
		\begin{equation}
			C_\nu\equiv {\sqrt{\pi}\over2}e^{{i\pi\nu\over2}+{i\pi\over4}}~~,~~C_\mu\equiv {\sqrt{\pi}\over2}e^{-{\pi\mu\over2}+{i\pi\over4}}
		\end{equation}

		Canonical quantisation of the scalar field is given by
		\begin{equation}
			\Phi({\bf x},\eta)=\int {d^dk\over (2\pi)^d} e^{i{\bf k}.{\bf x}} [a_{\bf k} F_\nu(k,\eta)+a^\dagger_{-\bf k} F_\nu(k,\eta)^*]
		\end{equation}
		where the annihilation operator $a_{\bf k}$ annihilates BD vacuum $\ket{0}$ of the scalar field. In the field eigenstate basis,
		\begin{equation}
			\Phi({\bf k},\eta)\ket{\varphi}=\varphi(-{\bf k},\eta)\ket{\varphi}
		\end{equation}
		the wavefunctional of the BD vacuum is represented as
		\begin{equation}
			\Psi[\varphi,\eta]=\langle{\varphi}|0\rangle= \exp[i\int {d^dk\over (2\pi)^d}\varphi({\bf k},\eta)\brf{\del_\eta F_\nu(k,\eta)^* \over 2(-\eta)^{d-1}F_\nu(k,\eta)^*}\varphi(-{\bf k},\eta)]
		\end{equation}
		
		In overdamped case, identification of source $\hat{\phi}$ of the dual CFT is given by
		\begin{equation}
			\label{overid}
			\varphi({\bf k},\eta)={f_{\nu}^*(k,\eta)\over f_{\nu}^*(k,\eta_B)}(-\eta_B)^{{d\over 2}-\nu}{\hat \phi}({\bf k})
		\end{equation} 
		where $f_\nu(k,\eta)$ is the late time behaviour of $F_\nu(k,\eta)$ given by
		\begin{equation}
			f_\nu(k,\eta)=(-\eta)^{d/2}[\alpha_\nu(-k\eta)^{-\nu}+\beta_\nu(-k\eta)^\nu]
		\end{equation}
		with
		\begin{equation}
			\alpha_\nu={2^{\nu-1}\Gamma[\nu]\over\sqrt{\pi}}e^{{i\pi\over2}(\nu-{1\over2})} ~~,~~ \beta_\nu=-{2^{-1-\nu}\sqrt{\pi}\over\Gamma[1+\nu]\sin[\pi\nu]}e^{-{i\pi\over2}(\nu+{1\over2})}
		\end{equation}
		Writing the wavefunctional in terms of the source ${\hat \phi}$, we get in the late time limit
		\begin{equation}
			\Psi[{\hat\phi},\eta_B]=\exp[-{i\over2}\int {d^dk\over (2\pi)^d}{\hat\phi}({\bf k})\brf{({d\over2}-\nu)(-\eta_B)^{-2\nu}+{\beta_\nu^*\over\alpha_\nu^*}2\nu k^{2\nu}}{\hat\phi}({-\bf k})]
		\end{equation}
		we see, by expressing the above wavefunctional in position space, that the finite part of the wavefunction coefficient determined by
		\begin{equation}
			\langle O({\bf x})O({\bf y}) \rangle={\delta^2 \log\Psi[{\hat\phi},\eta_B] \over \delta {\hat\phi}({\bf x})\delta {\hat\phi}({\bf y})}
		\end{equation}
		takes the form of CFT two point function of the primary fields with conformal dimension
		\begin{equation}
			\Delta_0={d\over2}+\nu
		\end{equation}
		
		In underdamped case, a similar identification of the source, as we mentioned above, cannot produce the wavefunction coefficient in the form of a CFT two point function. Note that, under an identification of source $J_+$ in the field eigenbasis given by
		\begin{equation}
			\label{J+relphi}
			\varphi({\bf k},\eta)=f_{\mu}^*(k,\eta)k^{i\mu}J_+({\bf k})
		\end{equation}
		where $f_\mu(k,\eta)$ is the late-time limit of $F_\mu(k,\eta)$ given by 
		\begin{equation}
			f_\mu(k,\eta)=(-\eta)^{d\over2}[\beta_\mu (-k\eta)^{-i\mu}+\alpha_\mu(-k\eta)^{i\mu}]
		\end{equation}
		with
		\begin{equation}
			\alpha_\mu=e^{-{\pi\mu\over2}+{i\pi\over4}}{\sqrt{\pi}(1+\coth[\pi\mu])\over 2^{1+i\mu}\Gamma[1+i\mu]} ~~,~~\beta_\mu=-e^{-{\pi\mu\over2}+{i\pi\over4}}{i2^{i\mu-1}\Gamma[i\mu]\over \sqrt{\pi}},
		\end{equation}
		the wavefunctional takes the form at late time limit as
		\begin{equation}
			\Psi[J_+]=\exp[-{i\over2}\int {d^dk\over(2\pi)^d} J_+({\bf k}) \brf{\alpha_\mu^*({d\over2}-i\mu)(-\eta_B)^{-2i\mu}+\beta_\mu^* ({d\over2}+i\mu)(-\eta_B)^{2i\mu}k^{4i\mu}+d\alpha_\mu^*\beta_\mu^* k^{2i\mu}} J_+(-{\bf k})]
		\end{equation}
		Note that, by expressing the above wavefunctional in position space, the wavefunction coefficient determined by
		\begin{equation}
			\langle O_+({\bf x})O_+({\bf y}) \rangle = {\delta^2 \log\Psi[J_+,\eta_B] \over \delta J_+({\bf x}) \delta J_+({\bf y})}
		\end{equation}
		has a finite non-local term in position space which is in the form of a CFT two point function with primary field of dimension $\Delta_{0+}$ given by
		\begin{equation}
			\label{dimsc+}
			\Delta_{0+}={d\over2}+i\mu
		\end{equation}
		Alongside the coefficient function also carries an unwieldy term which is both cut-off dependent and non-local in position space. It is found that this term is absent when the boundary source is identified representing the wavefunctional in the coherent state basis. More precisely, we define
		\begin{equation}
			{\tilde a}_{\bf k}=\sqrt{2\mu}k^{i\mu}[\alpha_\mu a_{\bf k}+\beta_\mu^* a_{-\bf k}^\dagger]
		\end{equation}
		and, in the eigenbasis of $\tilde a$ given by
		\begin{equation}
			{\tilde a}_{\bf k}\ket{\rho}=\rho({\bf k})\ket{\rho}
		\end{equation}
		one can get the BD wavefunctional at late time as follows,
		\begin{align}
			\Psi[\rho,\eta_B]&=\langle\rho|0\rangle=\int {D\varphi}\langle\rho|\varphi\rangle\langle\varphi|0\rangle \nonumber\\
			&= \exp[{1\over2}\int{d^dk\over (2\pi)^d}\rho^*({\bf k})\brf{(-\eta_B)^{-2i\mu}+{\beta_\mu^*\over \alpha_\mu^*}k^{2i\mu}}\rho^*(-{\bf k})] \label{2ndline}
		\end{align}
		where
		\begin{equation}
			\label{rhophiinp}
			\langle\rho|\varphi\rangle=\exp[\int{d^dk\over(2\pi)^d}{-i\over(\eta_B)^{d}}(i\sqrt{2\mu}(-\eta_B)^{{d\over2}-i\mu}\rho^*({\bf k})\varphi({\bf k},\eta_B)-{{d\over2}+i\mu \over2}\varphi({\bf k},\eta_B)\varphi(-{\bf k},\eta_B))]
		\end{equation}
		and, the equality at the second line of \eq{2ndline} is obtained at saddle point which relates $\varphi$ to $\rho$ as
		\begin{equation}
			\varphi={1\over\sqrt{2\mu}}[(-\eta)^{{d\over2}-i\mu}+{\beta_\mu^*\over\alpha_\mu^*}k^{2i\mu}(-\eta)^{{d\over2}+i\mu}]\rho^*(-{\bf k})
		\end{equation}
		Comparing with \eq{J+relphi}, we get
		\begin{equation}
			\rho^*({\bf k})=\sqrt{2\mu}\alpha_\mu^* J_+(-{\bf k})
		\end{equation}
		
		\subsection{Correction to Wavefunctional of Bunch-Davies vacuum}
		Introducing a scalar field $\phi$ of arbitrary mass $M_0$, the simplest scalar-Fermion interaction term can be written as
		\begin{equation}
			\label{sint}
			S_I=\lambda \int d^{d+1}x\sqrt{-g}\bar{\psi}{\phi}\psi
		\end{equation}
		which, by \eq{barpsisplit} can be written as
		\begin{equation}
			S_I=\lambda \int d^{d+1}x\sqrt{-g}({\psi}_+^\dagger{\phi}\psi_+ - {\psi}^\dagger_-{\phi}\psi_-)
		\end{equation}
		In presence of the interaction, Bunch Davies wave functional can be computed in path integral formalism as
		\begin{equation}
			\label{psiI}
			\Psi_I=\Psi \times \Psi_0\times e^{iS_I}
		\end{equation}
		where $\Psi$ is the wavefunctional of BD vacuum derived in the free Fermionic theory and $\Psi_0$ is that in the free scalar theory.
		Tree-level contribution in the correction term comes from the on-shell value of the action determined by the classical values of $\psi$ and $\phi$. On-shell solution is given by \eq{psi+cbd} and \eq{psi-cbd} for Fermionic field. For scalar field, it is given by
		\begin{equation}
			\phi({\bf k},\eta)=F_\nu^*(k,\eta)\phi_B({\bf k})
		\end{equation}
		Using these on-shell results, the correction term is calculated  at leading order of $\lambda$ as
		\begin{equation}
			\label{sicpm}
			iS_I=\brt{-i\lambda \int {d^dk_1d^dk_2d^dk_3\over (2\pi)^{3d}} \sum_{s_1,s_3} C_+^{s_1\dagger}({\bf k}_1)~(2\pi)^d\delta(-{\bf k}_1+{\bf k}_2+{\bf k}_3) I(k_1,k_2,k_3)\phi_B({\bf k}_2)C_-^{s_3}({\bf k}_3)}
		\end{equation}
		where $I(k_1,k_2,k_3)$ is given by
		\begin{align}
			I(k_1,k_2,k_3)\equiv \int_{-\infty}^{\eta_B}d\eta  F_\nu(k_2,\eta)^*&\left[e^{-2\pi M}H^1_{\nu_+}(-k_1\eta)H^2_{\nu_-}(-k_3\eta)\brf{{\gamma^ik_{1i}\over k_1}}\right.\nonumber\\
			&\left.-H^1_{\nu_-}(-k_1\eta)H^2_{\nu_+}(-k_3\eta)\brf{{\gamma^ik_{3i}\over k_3}}\right]
		\end{align}
		and, $C_\pm^s$ is given  by
		\begin{equation}
			C_\pm^s({\bf k})=C_4^s({\bf k})a_\pm^s({\bf k})
		\end{equation}
		Denoting 
		\begin{equation}
			\label{intdefk}
			\int_k\equiv \int {d^dk_1d^dk_2d^dk_3\over (2\pi)^{3d}}(2\pi)^d\delta(-{\bf k}_1+{\bf k}_2+{\bf k}_3)
		\end{equation}
		we write the result obtained in \eq{sicpm} in terms of $\psi_\pm$ and $\phi$ as,
		\begin{equation}
			\label{corbd0}
			iS_I=\brt{-\lambda\int_k\sum_s\psi_{+c}^{s_1\dagger}({\bf k}_1,\eta){e^{-i\pi\nu_+}\brf{{\gamma^ik_{1i}\over k_1}}I(k_1,k_2,k_3)\brf{{\gamma^ik_{3i}\over k_3}}\over (-\eta)^{1+{d}}H^1_{\nu_+}(-k_1\eta)H^2_{\nu_+}(-k_3\eta)F_\nu(k_2,\eta)^*}\phi({\bf k}_2,\eta)\psi_{-c}^{s_3}({\bf k}_3,\eta)}
		\end{equation}
		When exponentiated, the above result produces the correction factor in the wavefunctional of BD vacuum, see \eq{psiI}, at leading order to the coupling constant $\lambda$ due to inclusion of Yukawa interaction, \eq{sint}, at any time $\eta$. For algebraic simplicity in the further computations, We write the above expression as
		\begin{equation}
			\label{corbd}
			iS_I=\brt{-i\lambda\int_k\sum_s\psi_{+c}^{{s_1}\dagger}({\bf k}_1,\eta){J(k_1,k_2,k_3)\over (-\eta)^{1+{d}}H^1_{\nu_+}(-k_1\eta)H^2_{\nu_+}(-k_3\eta)F_\nu(k_2,\eta)^*}\phi({\bf k}_2,\eta)\psi_{-c}^{s_3}({\bf k}_3,\eta)}
		\end{equation}
		where
		\begin{align}
			J(k_1,k_2,k_3)\equiv \int_{-\infty}^{\eta_B}d\eta  F_\nu(k_2,\eta)^*&\left[e^{-\pi M}H^1_{\nu_+}(-k_1\eta)H^2_{\nu_-}(-k_3\eta)\brf{{\gamma^ik_{3i}\over k_3}}\right.\nonumber\\
			&\left.-e^{\pi M}H^1_{\nu_-}(-k_1\eta)H^2_{\nu_+}(-k_3\eta)\brf{{\gamma^ik_{1i}\over k_1}}\right] \label{Jdef}
		\end{align}
		Using the following identities
		\begin{equation}
			K_q(x)=
			\begin{cases}
				{\pi\over2}(-i)^{q+1}H^2_q(-ix)&{\rm if} -{\pi\over2}<arg(x)\leq \pi \\
				{\pi\over2}(i)^{q+1}H^1_q(ix)&{\rm if} -{\pi}<arg(x)\leq {\pi\over2}
			\end{cases}
			~~{\rm and}~~
			K_q(x)=K_{-q}(x)
		\end{equation}
		we can reduce \eq{Jdef} into
		\begin{align}
			J(k_1,k_2,k_3)= A\int_{i(-\eta_B)}^{i\infty}dz z^{d\over2} K_\nu(k_2z)&\left[K_{\nu_+}(-k_1z)K_{-\nu_-}(k_3z)\brf{{\gamma^ik_{3i}\over k_3}}\right.\nonumber\\
			&\left.-K_{-\nu_-}(-k_1z)K_{\nu_+}(k_3z)\brf{{\gamma^ik_{1i}\over k_1}}\right]\label{JdefK}
		\end{align}
		where we use $\eta=iz$, and the coefficient $A$ is given for overdamped and underdamped case as
		\begin{equation}
			A\equiv
			\begin{cases}
				{8\over\pi^3}C_\nu^* e^{{i\pi\over2}(\nu-d)}~, &{\rm overdamped} : \nu\in \real\\
				{8\over\pi^3}C_\mu^* e^{{i\pi\over2}(-i\mu-d)}~, &{\rm underdamped} : \mu\in\real
			\end{cases}		
		\end{equation}
		The $z$-integral is represented in the complex plane by the vertical blue line in the contour. Noting that at $z\rightarrow\infty$,
		\begin{align}
			&z^{d\over2} K_\nu(k_2z)\left[K_{\nu_+}(-k_1z)K_{-\nu_-}(k_3z)\brf{{\gamma^ik_{3i}\over k_3}}
			-K_{-\nu_-}(-k_1z)K_{\nu_+}(k_3z)\brf{{\gamma^ik_{1i}\over k_1}}\right]\nonumber\\
			&\rightarrow {z^{d-3\over2}\over\sqrt{-k_1k_2k_3}}\exp[(k_1-k_2-k_3)z]\brt{s_1\brf{{\gamma^ik_{3i}\over k_3}}-s_2\brf{{\gamma^ik_{1i}\over k_1}}} 
		\end{align}
		where $s_1,s_2$ are some coefficient, and as the momentum conservation implemented by $\delta(-{\bf k}_1+{\bf k}_2+{\bf k}_3)$ in the integral \eq{corbd}, see \eq{intdefk} alongside, implies $k_1\leq k_2+k_3$, 
		the integral on the arc (brown) at large radius in the contour vanishes\footnote{The equality $k_1=k_2+k_3$ is achieved for ${\bf k}_2\parallel{\bf k}_3$, which is very similar to OPE limit, a special case which we ignore drawing the conclusion.}. 
		The contribution from the arc (red) close to the origin $z=0$ is given by $\sim z^{{d\over2}-\nu}$ which also vanishes for both overdamped and underdamped cases.

		\begin{figure}[h]
			\centering
			\begin{tikzpicture}
				\draw (5,0)--(-5,0);
				\draw (0,5)--(0,-2);
				\node[text width=1.5cm] at (-0.1,0.5) {$-i\eta_B$};
				\node[text width=1.5cm] at (-1.5,2.5) {$J(k_1, k_2,k_3)$};
				\node[text width=1.5cm] at (.9,-0.2) {$-\eta_B$};
				\draw[blue] (4.9,0.05)--node{\midarrowL}(0.5,0.05);
				\draw[blue] (0.05,4.9)--node{\midarrowU}(0.05,0.5);
				\draw[brown] (4.9,0.05) node{\midarrowD} arc (0:90:4.85);
				\draw[red] (0.5,0.05) arc (0:90:0.45) node{\midarrowL};
			\end{tikzpicture}
			\caption{contour diagram to evaluate the integral ${J}(k_1, k_2, k_3)$.}
			\label{contour}
		\end{figure}
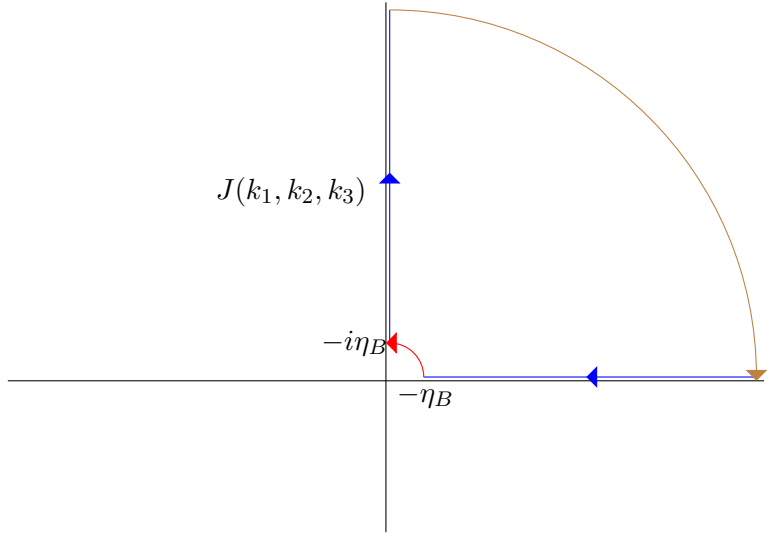

		Therefore, the integral \eq{JdefK} reduces to
		\begin{align}
			J(k_1,k_2,k_3)= A\int_{0}^{\infty}dz z^{d\over2} K_\nu(k_2z)&\left[K_{\nu_+}(-k_1z)K_{-\nu_-}(k_3z)\brf{{\gamma^ik_{3i}\over k_3}}\right.\nonumber\\
			&\left.-K_{-\nu_-}(-k_1z)K_{\nu_+}(k_3z)\brf{{\gamma^ik_{1i}\over k_1}}\right]\label{JdefK2}
		\end{align}

		\subsubsection{Overdamped scalar}
		\label{sec.yukover}
		In this subsection, we consider the scalar field in the interaction lies in the complementary series for which $\nu\in \real$. 
		At late time $\eta=\eta_B\rightarrow0$, we get from \eq{corbd}
		\begin{equation}
			\label{latetint}
			iS_I=\brt{-i\lambda \int_k\sum_s\psi_{+c}^{{s_1}\dagger}({\bf k}_1,\eta_B)J(k_1,k_2,k_3){C_{\nu,M}\times k_1^{\nu_+}{k_2}^{\nu}k_3^{\nu_+}\over (-\eta_B)^{{d-2iM}}(-\eta_B)^{{d\over2}-\nu}}\phi({\bf k}_2,\eta_B)\psi_{-c}^{s_3}({\bf k}_3,\eta_B)}
		\end{equation}
		where
		\begin{equation}
			C_{\nu,M}\equiv {\pi^{5/2}\over 2^{\nu+2iM}\Gamma[\nu_+]^2\Gamma[\nu]}e^{{i\pi\nu/2}-{i\pi/4}}
		\end{equation}
		Expressing the correction \eq{latetint} in terms of sources identified in \eq{rhopm_s} and \eq{overid} as
		\begin{equation}
			\delta(\log \Psi)=\brt{-i\lambda\int_k\sum_s\rho_{+}^{{s_1}\dagger}({\bf k}_1)[{C_{\nu,M} J(k_1,k_2,k_3) k_1^{\nu_+}{k_2}^{\nu}k_3^{\nu_+}}]\hat{\phi}({\bf k}_2)\rho_{-}^{s_3}({\bf k}_3)}
		\end{equation}
		the wave function coefficient due to the sources $\hat\phi$ and $\rho_\pm$ is found to be
		\begin{equation}
			\label{3ptmom}
			\langle{\hat\zeta}_{+}({\bf k}_1){\mathbb O}({\bf k}_2){\hat\zeta}_{-}^\dagger({\bf k}_3)\rangle=i\lambda {C_{\nu,M} J_{+-}(k_1,k_2,k_3) k_1^{\nu_+}{k_2}^{\nu}k_3^{\nu_+}}\delta(-{\bf k}_1+{\bf k}_2+{\bf k}_3)
		\end{equation}
		where $J_{+-}(k_1,k_2,k_3)$ is given by the top-right quarter block of \eq{JdefK2},
		 	\begin{align}
		 	J_{+-}(k_1,k_2,k_3)= A\int_{0}^{\infty}dz z^{d\over2} K_\nu(k_2z)&\left[K_{\nu_+}(-k_1z)K_{-\nu_-}(k_3z)\brf{{\gamma_{+-}^ik_{3i}\over k_3}}\right.\nonumber\\
		 	&\left.-K_{-\nu_-}(-k_1z)K_{\nu_+}(k_3z)\brf{{\gamma_{+-}^ik_{1i}\over k_1}}\right]\label{J+-}
		 \end{align}

		Now, we are interested in the position space representation of the above expression we obtained for wavefunction coefficient given by
		\begin{equation}
			\langle{\hat\zeta}_{+}({\bf x}_1){\mathbb O}({\bf x}_2){\hat\zeta}_{-}^\dagger({\bf x}_3)\rangle=\int \brf{\prod_{p=1}^{3}{d^dk_p\over (2\pi)^d}} \langle{\hat\zeta}_{+}({\bf k}_1){\mathbb O}({\bf k}_2){\hat\zeta}_{-}^\dagger({\bf k}_3)\rangle \exp[{i (-{\bf k}_1.{\bf x}_1+{\bf k}_2.{\bf x}_2+{\bf k}_3.{\bf x}_3)}]
		\end{equation}
		Using \eq{3ptmom}, we get
		\begin{align}
			&\langle{\hat\zeta}_{+}({\bf x}_1){\mathbb O}({\bf x}_2){\hat\zeta}_{-}^\dagger({\bf x}_3)\rangle =i\lambda C_{\nu,M}\int_k J_{+-}(k_1,k_2,k_3) k_1^{\nu_+}{k_2}^{\nu}k_3^{\nu_+} \exp[{i (-{\bf k}_1.{\bf x}_1+{\bf k}_2.{\bf x}_2+{\bf k}_3.{\bf x}_3)}]\nonumber\\	
			&=i\lambda C_{\nu,M}\int d^d x\int \prod_{p=1}^{3}{d^dk_p\over(2\pi)^d} J_{+-}(k_1,k_2,k_3) k_1^{\nu_+}{k_2}^{\nu}k_3^{\nu_+} \exp[{-i (-{\bf k}_1.({\bf x}-{\bf x}_1)+{\bf k}_2.({\bf x}-{\bf x}_2)+{\bf k}_3.({\bf x}-{\bf x}_3))}] \label{x-xp}
		\end{align}
		Using \eq{JdefK2} and the following integral identity
		\begin{equation}
			\int {d^dk\over(2\pi)^d} k^\nu K_\nu(kz)e^{i{\bf k}.{\bf r}}={\Gamma[{d\over2}+\nu]\over 2^{1-\nu}\pi^{d/2}}z^{-{d\over2}}\brf{z\over z^2+r^2}^{{d\over 2}+\nu}
		\end{equation}
		\eq{x-xp} is reduced to 
		\begin{align}
			\langle{\hat\zeta}_{+}({\bf x}_1)&{\mathbb O}({\bf x}_2){\hat\zeta}_{-}^\dagger({\bf x}_3)\rangle=-\lambda C_{\nu,M}{\Gamma[\Delta_0]\Gamma[{d+1\over2}+iM]\Gamma[{d-1\over 2}+iM]\over 2^{3-\nu-2iM}\pi^{3d/2}}e^{-i\pi\nu_-}A\times\nonumber\\&\gamma_{+-}^i\int d^d x \int_{0}^{\infty}{dz\over z^d}\brt{\del_i^{{\bf x}_3}h_{\nu_+}(r_1)h_{\nu}(r_2)h_{-\nu_-}(r_3)-\del_i^{{\bf x}_1}h_{-\nu_-}(r_1)h_{\nu}(r_2)h_{\nu_+}(r_3)} \label{int123}
		\end{align}
		where
		\begin{equation}
			h_\nu(r)=\brf{z\over z^2+r^2}^{{d\over 2}+\nu} ~~,~~ {\bf r}_p={\bf x}-{\bf x}_p
		\end{equation}
		Some algebraic computations yield the integral in \eq{int123} to
		\begin{align}
			&\gamma_{+-}^i\int d^d x \int_{0}^{\infty}{dz\over z^d}\brt{\del_i^{{\bf x}_3}h_{\nu_+}(r_1)h_{\nu}(r_2)h_{-\nu_-}(r_3)-\del_i^{{\bf x}_1}h_{-\nu_-}(r_1)h_{\nu}(r_2)h_{\nu_+}(r_3)}\nonumber\\
			=&\gamma_{+-}^i(x_1-x_3)_i\int d^d x \int_{0}^{\infty}{dz\over z^{d+1}}[h_{\nu_+}(r_1)h_{\nu}(r_2)h_{\nu_+}(r_3)] (d-2\nu_-)
		\end{align}
		Note that the similar integral obtained above has been computed in Appendix (C.2) of \cite{roy_dscft}, see also \cite{Freedman:1998tz}. Here we therefore directly present the value of the above integral as follows
		\begin{align}
			\int d^d x \int_{0}^{\infty}{dz\over z^{d+1}}[h_{\nu_+}(r_1)h_{\nu}(r_2)h_{\nu_+}(r_3)]={B[\nu,\nu_+]\over |{\bf x}_1-{\bf x}_2|^{\Delta_0}|{\bf x}_2-{\bf x}_3|^{\Delta_0}|{\bf x}_3-{\bf x}_1|^{1+2\Delta_+ -\Delta_0}}
		\end{align}
		where
		\begin{equation}
			\Delta_0={d\over 2}+\nu
		\end{equation}
		is the conformal dimension of the scalar field and $\Delta_+$ is given in \eq{Fdim}. $B[\nu,\nu_+]$ is given by
		\begin{equation}
			B[\nu,\nu_+]={\pi^{d/2}\over2}{\Gamma[{2\nu_+ +\Delta_0 \over2}]\Gamma[{\Delta_0\over2}]^2\Gamma[{2\nu_++d-\Delta_0\over2}]\over\Gamma[{d\over2}+\nu_+]^2\Gamma[\Delta_0]}
		\end{equation}
		Using the above results, \eq{int123} is given as
		\begin{align}
			\langle{\hat\zeta}_{+}({\bf x}_1){\mathbb O}({\bf x}_2){\hat\zeta}_{-}^\dagger({\bf x}_3)\rangle=&-\lambda C_{\nu,M}{\Gamma[\Delta_0]\Gamma[{d+1\over2}+iM]\Gamma[{d-1\over 2}+iM](d-2\nu_-)\over 2^{3-\nu-2iM}\pi^{3d/2}}e^{-i\pi\nu_-}A\times\nonumber\\&{\gamma_{+-}^i(x_1-x_3)_i\over |{\bf x}_3-{\bf x}_1|}{B[\nu,\nu_+]\over |{\bf x}_1-{\bf x}_2|^{\Delta_0}|{\bf x}_2-{\bf x}_3|^{\Delta_0}|{\bf x}_3-{\bf x}_1|^{2\Delta_+ -\Delta_0}}\label{3ptpos}
		\end{align}
		which is indeed in the form of the three point correlation function of a CFT with scalar field of dimension $\Delta_0$ and Fermionic fields ${\hat\zeta}_+,{\hat\zeta}_-$ of dimensions $\Delta_+,\Delta_-$, see Appendix B of \cite{Dobrev:1973zg}, as promised in Section \ref{sec.holo}. 
		
		It is interesting at this point to discuss two OPE limits of the above expression. First, we take the scalar field ${\mathbb O}({\bf x}_2)$ very close to one of the Fermionic insertions, say ${\hat\zeta}_-^\dagger({\bf x}_3)$, i.e., $|{\bf x}_2 - {\bf x}_3|\equiv \epsilon_{23}\ll |{\bf x}_1-{\bf x}_2|$. In this limit, \eq{3ptpos} reduces to
		\begin{align}
			\lim_{{\bf x}_2\rightarrow {\bf x}_3}	\langle{\hat\zeta}_{+}({\bf x}_1){\mathbb O}({\bf x}_2){\hat\zeta}_{-}^\dagger({\bf x}_3)\rangle=-{\lambda\over\epsilon_{23}^{\Delta_0}} {\gamma_{+-}^i(x_1-x_2)_i\over |{\bf x}_1-{\bf x}_2|}{Q_{\nu,,M}\over |{\bf x}_1-{\bf x}_2|^{2\Delta_+}}\label{3ptope1}
		\end{align}
		where
		\begin{equation}
			Q_{\nu,M} \equiv AC_{\nu,M}{\Gamma[\Delta_0]\Gamma[{d+1\over2}+iM]\Gamma[{d-1\over 2}+iM]\over 2^{3-\nu-2iM}\pi^{3d/2}}e^{-i\pi\nu_-}{B[\nu,\nu_+] } (d-2\nu_-)
		\end{equation}
		Note that, the limit reduces the three point correlation is the form a CFT two point function between two Fermionic operator with conformal dimension $\Delta_+$, as mentioned in Section \ref{Sec.discuss}. Second, we take the Fermionic insertions closer to each other, i.e., $|{\bf x}_1 - {\bf x}_3|\equiv \epsilon_{13}\ll |{\bf x}_1-{\bf x}_2|$. In this limit, \eq{3ptpos} reduces to
		\begin{align}
			\lim_{{\bf x}_1\rightarrow {\bf x}_3}	\langle{\hat\zeta}_{+}({\bf x}_1){\mathbb O}({\bf x}_2){\hat\zeta}_{-}^\dagger({\bf x}_3)\rangle=-{\lambda\over\epsilon_{13}^{2\Delta_+-\Delta_0}} {Q_{\nu,,M}\over |{\bf x}_1-{\bf x}_2|^{2\Delta_0}}{(\gamma_{+-}^i{\hat n}_i)}\label{3ptope2}
		\end{align}
		where ${\hat n}^i\equiv {\epsilon_{13}^i \over \epsilon_{13}}$ is the unit vector component in the direction of the infinitesimal vector $\epsilon_{13}^i\equiv ({\bf x}_1-{\bf x}_3)^i$. Note that \eq{3ptope2} is in the form of a CFT two point function between scalar fields of dimension $\Delta_0$, up to the `polarisation factor' $\gamma_{+-}^i{\hat n}_i$.

		\subsubsection{Underdamped Scalar}
		In this subsection, we consider the scalar field in the interaction lies in the principal series for which $\nu=i\mu$ where $\mu\in\real$. In this case, we first write \eq{corbd}, at the late-time limit, in terms of $J_+$ and $\rho_\pm$, using \eq{J+relphi} and \eq{rhopm_s}, as
		\begin{equation}
			iS_I=-i\lambda{\pi^2\over2^{2\nu_+}\Gamma[\nu_+]^2}\int_k\sum_s\rho_{+}^{{s_1}\dagger}({\bf k}_1)J(k_1,k_2,k_3){k_1^{\nu_+}k_3^{\nu_+}}k_2^{i\mu}J_+({\bf k}_2)\rho_{-}^{s_3}({\bf k}_3)
		\end{equation}
		Putting above expression in \eq{psiI}, one can thereby obtain the wavefunctional in terms of $J_+$ and $\rho_\pm$. The coherent state basis representation of the resulting wavefunctional is given by
		\begin{equation}
			\label{psiIrho}
			\Psi_I[\rho,\rho_+^\dagger,\rho_-]=\int {D J_+} \Psi_\rho^*[J_+]\Psi_I[J_+,\rho_+^\dagger,\rho_-]
		\end{equation}
		In the above integrand, $\Psi_\rho^*[J_+]$ is the value of $\langle\rho|\varphi\rangle$, \eq{rhophiinp}, evaluated in terms of $J_+$ using \eq{J+relphi} as,
		\begin{empheq}{multline}
			\label{formfa}
			\Psi_\rho^*[J_+]=\exp\left[\int {d^d{\bf k}\over(2\pi)^d}{-i\over(-\eta)^d}\left\{i\sqrt{2\mu} f_\mu^\ast(k,\eta) k^{i\mu}(-\eta)^{{d\over2}-i\mu}\rho^*({\bf k}) J_+({\bf k})\right.\right.\\
			\left.\left.-{\Delta_{0+}\over2}(f_\mu^\ast(k,\eta))^2 k^{2i\mu} J_+({\bf k})J_+(-{\bf k})\right\}\right]
		\end{empheq}
		And, $\Psi_I[J_+,\rho_+^\dagger,\rho_-]$ is given in \eq{psiI}. All together, at late-time,
		\begin{align}
			\Psi_I[\rho,\rho_+^\dagger,\rho_-]=\Psi[\rho_+^\dagger,\rho_-]\int {D J_+}\exp\left[\int \frac{d^d\mathbf{k}}{(2\pi)^d}W_1\rho^{\ast}({\bf k}) J_+({\bf k})+i\int \frac{d^d\mathbf{k}}{(2\pi)^d} ~W_2 J_+({\bf k})J_+(-{\bf k})\right.\nonumber\\
			\left.+{i\lambda}\int_k\sum_s \rho_+^{s_1\dagger}({\bf k}_1) Q_2J_{+}({\bf k_2})\rho_-^{s_3}({\bf k}_3)\right] 
		\end{align}
		where
		\begin{equation}
			\label{W1W2}
			W_1=\sqrt{2\mu} f_\mu^\ast(k,\eta) k^{i\mu}(-\eta)^{-\Delta_{0+}} ~~,~~W_2=k^{2i\mu}\brf{\frac{f_\mu^\ast(k,\eta)\partial_\eta f_\mu^\ast(k,\eta)}{2(-\eta)^{d-1}}+{\Delta_{0+}(f_\mu^\ast(k,\eta))^2 \over2(-\eta)^d}}
		\end{equation}
		and,
		\begin{equation}
			Q_2=-{\pi^2\over2^{2\nu_+}\Gamma[\nu_+]^2} J(k_1,k_2,k_3){k_1^{\nu_+}k_3^{\nu_+}}k_2^{i\mu}
		\end{equation}
		In semi-classical regime, the integral \eq{psiIrho} is computed at the saddle point approximation. The saddle point is found by extremising the integrand as,
		\begin{equation}
			W_1\rho^*({\bf k})+2iW_2 J_+(-{\bf k})+i\lambda\int {d^dk_1d^dk_3\over (2\pi)^{d}}\delta(-{\bf k}_1+{\bf k}+{\bf k}_3)\sum_s\rho_+^{s_1\dagger}({\bf k}_1)Q_2\rho_-^{s_3}({\bf k}_3)=0
		\end{equation}
		which yields
		\begin{equation}
			J_+({-\bf k})={i\over 2W_2}[W_1\rho^*({\bf k})+i\lambda\int {d^dk_1d^dk_3\over (2\pi)^{d}}\delta(-{\bf k}_1+{\bf k}+{\bf k}_3)\sum_s\rho_+^{s_1\dagger}({\bf k}_1)Q_2\rho_-^{s_3}({\bf k}_3)]
		\end{equation}
		The $O(\lambda)$ correction to the wavefunctional is therefore given by
		\begin{align}
			\delta\log[\Psi_I[\rho,\rho_+^\dagger,\rho_-]]=\int \frac{d^d\mathbf{k}}{(2\pi)^d}W_1\rho^{\ast}({\bf k}) \delta J_+({\bf k})+i\int \frac{d^d\mathbf{k}}{(2\pi)^d} ~W_2 (\delta J_+({\bf k})J_+^{(0)}(-{\bf k})+ J_+^{(0)}({\bf k})\delta J_+(-{\bf k}))\nonumber\\
			+{i\lambda}\int_k \sum_s\rho_+^{s_1\dagger}({\bf k}_1) Q_2J_{+}^{(0)}({\bf k_2})\rho_-^{s_3}({\bf k}_3) \label{corund}
		\end{align}
		where
		\begin{equation}
			\label{delJ0J}
			\delta J_+ (-{\bf k})=-{\lambda\over2W_2}\int {d^dk_1d^dk_3\over (2\pi)^{d}}\delta(-{\bf k}_1+{\bf k}+{\bf k}_3)\sum_s\rho_+^{s_1\dagger}({\bf k}_1)Q_2\rho_-^{s_3}({\bf k}_3)~~,~~J^{(0)}_+(-{\bf k})={iW_1\over 2W_2}\rho^*({\bf k})
		\end{equation}
		One can show, by putting \eq{delJ0J} in \eq{corund}, that the first two terms in \eq{corund} cancel each other yielding
		\begin{align}
			\delta\log[\Psi_I[\rho,\rho_+^\dagger,\rho_-]]=
			{-\lambda}\int_k {W_1\over 2W_2}\sum_s\rho_+^{s_1\dagger}({\bf k}_1) Q_2\rho^*(-{\bf k}_2)\rho_-^{s_3}({\bf k}_3)
		\end{align}
		Using \eq{W1W2}, 
		\begin{align}
			\delta\log[\Psi_I[\rho,\rho_+^\dagger,\rho_-]]=
			{-i\lambda}C_{\mu,M}\int_k \sum_s\rho_+^{s_1\dagger}({\bf k}_1)  J(k_1,k_2,k_3){k_1^{\nu_+}k_3^{\nu_+}}k_2^{i\mu}\rho^*(-{\bf k}_2)\rho_-^{s_3}({\bf k}_3)
		\end{align}
		where
		\begin{equation}
			C_{\mu,M}\equiv{\pi^2\over2^{2\nu_+}\Gamma[\nu_+]^2}{1\over \sqrt{2\mu}\alpha_\mu^*}
		\end{equation}
		Noting that $\rho({\bf k})=\int d^dx \rho({\bf x}) e^{-i{\bf k}.{\bf x}}$, in position space,
		\begin{align}
			\delta\log[\Psi_I[\rho,\rho_+^\dagger,\rho_-]]=
			\int ({\prod_{p=1}^3}d^dx_p) \sum_s{\hat\rho}_+^{s_1\dagger}({\bf x}_1) {\langle {\hat\zeta}_{+}({\bf x}_1) O_+({\bf x}_2) {\hat\zeta}_{-}^\dagger({\bf x}_3) \rangle } \rho^*({\bf x}_2){\hat\rho}_-^{s_3}({\bf x}_3)
		\end{align}
		where,
		\begin{align}
			{\langle {\hat\zeta}_{+}({\bf x}_1) O_+({\bf x}_2) {\hat\zeta}_{-}^\dagger({\bf x}_3) \rangle }= &{i\lambda}C_{\mu, M}	\int d^dx\int {\prod_{p=1}^3 {d^dk_p\over (2\pi)^d}} J_{+-}(k_1,k_2,k_3){k_1^{\nu_+}k_2^{i\mu}k_3^{\nu_+}}\nonumber\\ &\times\exp[{-i (-{\bf k}_1.({\bf x}-{\bf x}_1)+{\bf k}_2.({\bf x}-{\bf x}_2)+{\bf k}_3.({\bf x}-{\bf x}_3))}]
		\end{align}
		The above integral is evaluated similarly to the above subsection as,
		\begin{align}
			{\langle {\hat\zeta}_{+}({\bf x}_1) O_+({\bf x}_2) {\hat\zeta}_{-}^\dagger({\bf x}_3) \rangle }=& {-\lambda}C_{\mu, M}{\Gamma[\Delta_{0+}]\Gamma[{d+1\over2}+iM]\Gamma[{d-1\over 2}+iM](d-2\nu_-)\over 2^{3-i\mu-2iM}\pi^{3d/2}}e^{-i\pi\nu_-}A\times\nonumber\\&{\gamma_{+-}^i(x_1-x_3)_i\over |{\bf x}_3-{\bf x}_1|}{B[i\mu,\nu_+]\over |{\bf x}_1-{\bf x}_2|^{\Delta_{0+}}|{\bf x}_2-{\bf x}_3|^{\Delta_{0+}}|{\bf x}_3-{\bf x}_1|^{2\Delta_+ -\Delta_{0+}}}
		\end{align}  
		which is indeed in the form of the three point correlation function of a CFT with scalar field of dimension $\Delta_{0+}$ and Fermionic field of dimension $\Delta_{+}$. 
		Also note that similar comments hold true for the OPE limits as discussed in the above subsection \ref{sec.yukover}.

		\newpage
		\bibliographystyle{JHEP}
		\bibliography{dSCFT_fermion}

	\end{document}

%% file: packdefs.tex
\usepackage{setspace}
\usepackage{graphicx}
\usepackage{wrapfig} 
\usepackage{caption}
\usepackage{subcaption}
\usepackage{hyperref}
\usepackage{tikz} 
\usepackage{empheq}
\usepackage{physics}
\usepackage{amsmath}
\usepackage{mathdots}
\usepackage{yhmath}
\usepackage{cancel}
\usepackage{color}
\usepackage{siunitx}
\usepackage{array}
\usepackage{multirow}
\usepackage{amssymb}
\usepackage{gensymb}
\usepackage{tabularx}
\usepackage{extarrows}
\usepackage{booktabs}
\usepackage{mathtools}
\usepackage{simplewick}
\usepackage[obeyFinal]{todonotes}

\usetikzlibrary{fadings}
\usetikzlibrary{patterns}
\usetikzlibrary{shadows.blur}
\usetikzlibrary{shapes}
\usetikzlibrary{arrows}

\definecolor{refkey}{gray}{0.45}
\definecolor{labelkey}{RGB}{155,48,48} 
\definecolor{UI_blue}{RGB}{32, 64, 151}
\definecolor{UI_red}{RGB}{187, 62, 24}
\definecolor{UI_blue2}{RGB}{0, 84, 147}
\definecolor{UI_red2}{RGB}{159, 32, 66}
\definecolor{UI_gray}{RGB}{169, 169, 169}
\definecolor{UI_sepia}{RGB}{112, 66, 20}
\definecolor{UI_bittersweet}{RGB}{254, 111, 94}
\definecolor{UI_emerald}{RGB}{80, 200, 120}
\definecolor{UI_olivegreen}{RGB}{181, 179, 92}
\definecolor{UI_cadetblue}{RGB}{95, 158, 160}
\definecolor{UI_fuchsia}{RGB}{255, 0, 255}
\definecolor{UI_midnightblue}{RGB}{25, 25, 112}
\definecolor{UI_royalblue}{RGB}{0,35, 102}
\definecolor{UI_periwinkle}{RGB}{204, 204, 255}
\definecolor{UI_redorange}{RGB}{255, 83, 73}
\definecolor{UI_brickred}{RGB}{203,65,84}	
\definecolor{UI_forestgreen}{RGB}{34, 139, 34}
\definecolor{UI_tan}{RGB}{210,180,140}	
\definecolor{UI_burlywood}{RGB}{222,184,135}
\definecolor{UI_burlywood}{RGB}{192,64,0}
\definecolor{UI_darkorchid}{RGB}{153,50,204}

\newcommand{\midarrowD}{\tikz \draw[-triangle 90] (0,0) -- +(0,-0.1);}
\newcommand{\midarrowU}{\tikz \draw[-triangle 90] (0,0) -- +(0,0.1);}
\newcommand{\midarrowL}{\tikz \draw[-triangle 90] (0,0) -- +(-0.1,0.0);}

\newcommand{\eq}[1]{eq.(\ref{#1})}
\newcommand{\brf}[1]{\left(#1\right)}
\newcommand{\brs}[1]{\left\{#1\right\}}
\newcommand{\brt}[1]{\left[#1\right]}

\def\beq{\begin{eqnarray}}\def\eeq{\end{eqnarray}}
\def\be{\begin{equation}}\def\ee{\end{equation}}

\def\mes[#1]{d^{3}{#1}}

\newcommand{\fint}[1]{\int {d^dk_{#1}\over (2\pi)^d}}
\def\del{\partial}

\def\del{\partial}

\reversemarginpar